\documentclass[twocolumn,floatfix,superscriptaddress,a4paper,showkeys,nofootinbib,reprint]{revtex4-1}
\usepackage{epsfig}
\usepackage{latexsym}
\usepackage{xspace}
\usepackage[colorlinks=true,linktocpage=true,linkcolor=blue,citecolor=blue,allcolors=blue]{hyperref}
\usepackage[utf8]{inputenc}
\usepackage{indentfirst}
\usepackage{enumerate}
\usepackage{listings}
\usepackage{color}

\usepackage[caption=false,position=top]{subfig}

\usepackage{amsmath}
\usepackage{amssymb}
\usepackage[english]{babel}
\usepackage{url}

\newcommand{\cc}[1]{\chi_{#1}^B}

\newcommand{\eq}[1]{\begin{align} #1 \end{align}}
\newcommand{\mean}[1]{\langle #1 \rangle}

\newcommand{\sNN}{\sqrt{s_{\rm NN}}}
\newcommand{\Nacc}{N_{\rm acc}}

\usepackage[normalem]{ulem}  

\begin{document}

\title{Canonical statistical hadronization with local baryon conservation for higher-order cumulants}

\author{Mario Ciacco}
    \affiliation{Dipartimento di Fisica dell’Università and Sezione INFN, Turin, Italy}

\author{Sourav Kundu}
    \affiliation{Experimental Physics Department, CERN, CH-1211 Geneve 23, Switzerland}

\author{Volodymyr~A.~Kuznietsov}
    \affiliation{Physics Department, University of Houston, 3507 Cullen Blvd, Houston, TX 77204, USA}
    \affiliation{Bogolyubov Institute for Theoretical Physics, 03680 Kyiv, Ukraine}

\author{Maximiliano Puccio}
    \affiliation{Experimental Physics Department, CERN, CH-1211 Geneve 23, Switzerland}
    
\author{Volodymyr~Vovchenko}
    \affiliation{Physics Department, University of Houston, 3507 Cullen Blvd, Houston, TX 77204, USA}

\begin{abstract}
We study higher-order cumulants of the conserved baryon number at the LHC within the canonical ensemble with local baryon conservation.
We generalize the density correlations approach of [V.~Vovchenko, \href{https://doi.org/10.1103/PhysRevC.110.L061902}{{Phys. Rev. C {\bf 110}, L061902 (2024)}}] to incorporate the effect of Gaussian local conservation in spatial rapidity space in cumulants up to 6th order.
Gaussian local conservation improves upon the commonly employed $V_c$ approach, yielding comparable predictions at midrapidity, but marked differences for larger rapidity acceptances.
Our coordinate-space results are in exact agreement with the diffusion master equation approach for all cumulant ratios up to $\kappa_6/\kappa_2$.
Using the blast-wave model to apply kinematic cuts, we obtain predictions for net-proton cumulants in O--O and Pb--Pb collisions at the LHC that establish an ideal hadron gas baseline.
We find that local baryon conservation alone can drive $\kappa_6/\kappa_2$ to small or even negative values in restricted acceptance, a behavior often associated with chiral criticality.
The conservation baseline must therefore be carefully accounted for when interpreting upcoming LHC measurements.
\end{abstract}


\maketitle

\section{Introduction}
\label{sec-intro}

Fluctuations of conserved charges encode the QCD medium response to variations of charge density conditions in equilibrium and provide information about the nature of the medium.
In particular, fluctuations are sensitive to the phase structure, and measurements of net-proton cumulants are actively employed in the search for the QCD critical point at finite baryon density~\cite{Stephanov:1999zu,Bzdak:2019pkr}, with data collected at RHIC, SPS, and the future FAIR facility.
The high-energy collisions at the LHC probe approximately net baryon-free matter~($\mu_B = 0$), where lattice QCD calculations predict a smooth chiral crossover transition, characterized by a pseudocritical temperature of $T_{\rm pc} \approx 155$~MeV~\cite{HotQCD:2018pds,Borsanyi:2020fev}.
The chiral crossover is a reflection of chiral symmetry restoration and the corresponding chiral phase transition in the limit of vanishing quark masses~\cite{Pisarski:1983ms,HotQCD:2019xnw}.
Remnants of chiral criticality are thought to be encoded in higher-order baryon number susceptibilities. In particular, the chiral transition is characterized by a drop in the $\chi_4^B/\chi_2^B$ ratio from unity, while a negative sign of $\chi_6^B$ is thought to reflect chiral criticality~\cite{Friman:2011pf}.

Thermal model analyses of mean hadron yields indicate that the hadrochemical freeze-out at the LHC takes place at temperatures very similar to $T_{\rm pc}$~\cite{ALICE:2022wpn,Andronic:2017pug,Vovchenko:2018fmh}.
It is therefore natural to ask whether the measurements of fluctuations, particularly the net-proton cumulants, are sensitive to chiral criticality and can potentially provide a direct experimental signature of the chiral crossover transition.
Such an analysis requires precise control over effects not related to equilibrium susceptibilities. 

Perhaps the most important among such effects is the canonical suppression due to (local) baryon conservation, which reduces fluctuations relative to the grand-canonical baseline~\cite{Begun:2004gs,Bzdak:2012an}.
This mechanism appears to be the primary driver behind the qualitative behavior of second-order (net-)proton cumulants measured at RHIC~\cite{Braun-Munzinger:2020jbk,Vovchenko:2021kxx} and the LHC~\cite{Vovchenko:2020kwg}, which show a mild suppression of fluctuations relative to the Poisson statistics baseline of unity.

The canonical effect of baryon conservation reflects the fact that the total baryon number is conserved over the course of a heavy-ion collision and does not fluctuate. 
Thus, cumulants $\kappa_n$ of baryon number $B_{4\pi}$ in full space vanish regardless of the equation of state and the corresponding equilibrium grand-canonical susceptibilities.
Measurements of cumulants in a restricted phase space can mimic the grand-canonical ensemble under certain conditions, namely if the acceptance region is simultaneously large relative to the correlation length and small relative to the full volume~\cite{Koch:2008ia}.
In practice, such conditions are never truly realized in heavy-ion collisions~\cite{Vovchenko:2020tsr,Koch:2025cog}.
In the simplest case, one models the effect of baryon conservation by considering the canonical ensemble description of the whole fireball~\cite{Begun:2004gs,Bzdak:2012an,Vovchenko:2020tsr}.
This global constraint, typically implemented through a Kronecker delta enforced on the global net baryon number, leads to (anti)correlations among baryon charges.
While it enforces exact global conservation, it largely ignores the dynamical nature of heavy-ion collisions.
In particular, the global constraint correlates baryon charges spread over the entire fireball, including forward and backward rapidity regions that are causally disconnected, unless the production and equilibration of baryon charge took place at very early times of the collision~\cite{Castorina:2013mba}.
This led to the concept of local charge conservation, typically in rapidity space, where the correlations induced by exact charge conservation are spread over a localized region.

There have been several approaches in the literature to implement local charge conservation.
Reference~\cite{Bozek:2012en} implemented it by requiring each particle to be balanced by an antiparticle at the same coordinate at freeze-out.
This models the extreme case of maximally-local charge conservation but does not provide an easy way to describe an intermediate case where charge correlations are spread over a finite range.
Reference~\cite{Sakaida:2014pya} considered local production of balancing charged hadrons followed by diffusive dynamics in (spatial) rapidity space, leading to the spread of canonical correlations over a non-zero rapidity range, depending on the diffusion coefficient and the duration of the diffusive stage.
This approach offered insights into the behavior of net-charge cumulants up to the fourth order but was restricted to coordinate space.
Reference~\cite{Braun-Munzinger:2023gsd} introduced local baryon number conservation by combining the sampling of full-space multiplicities from a canonical ensemble distribution with a correlated sampling of rapidities for all baryon-antibaryon pairs.
When applied to second-order net-baryon cumulants, the procedure yields a characteristic suppression consistent with the diffusion model results of~\cite{Sakaida:2014pya}.
The approach of~\cite{Braun-Munzinger:2023gsd} allows both sampling from arbitrary rapidity distributions and the possibility of replacing (anti)correlations among unlike baryons by (anti)correlations among like baryons~($BB$ and $\bar{B}\bar{B}$).

Finally, a canonical statistical model with local charge conservation was developed in Refs.~\cite{Vovchenko:2018fiy,Vovchenko:2019kes}, originally for describing canonical suppression of the yields of hadrons and light nuclei in small systems at the LHC.
This approach assumes that the effective volume $V_c$ across which the charge is conserved exactly is smaller than the full volume, and it is localized to $k$ units around midrapidity, $V_c = k dV/dy$.
Performing Monte Carlo sampling of particle momenta from a canonical-ensemble hadron resonance gas in volume $V_c$ combined with a realistic blast-wave flow profile, as implemented in \texttt{Thermal-FIST}~\cite{Vovchenko:2019pjl}, allowed one to study the effects of local charge conservation on correlations and fluctuations of various hadron numbers~\cite{ALICE:2022xpf,ALICE:2022xiu,ALICE:2024rnr}.

One limitation of the $V_c$ approach is that it assumes a box-like correlation due to local charge conservation in rapidity and neglects hadrons at forward and backward rapidities, $|\eta_s| > k/2$.
This issue was addressed in Ref.~\cite{Vovchenko:2024pvk}, where local charge conservation was introduced via a Gaussian kernel in rapidity.
The analysis was restricted to second-order cumulants, constraining the range of local baryon conservation in $\sNN = 5.02$ TeV Pb-Pb collisions to $\sigma_\eta \approx 0.78$ rapidity units.
In Ref.~\cite{Parra:2025fse}, the formalism was applied to the analysis of net-charge fluctuations, obtaining moderate evidence for the freeze-out of charge fluctuations in the QGP.

In this work we generalize the approach of~\cite{Vovchenko:2024pvk} to higher-order cumulants.
The three main new elements are: (i)~finite-volume-consistent $n$-point Gaussian conservation kernels with reflecting boundary conditions, (ii)~closed-form acceptance formulas for net-baryon and species-decomposed cumulants up to sixth order, and (iii)~realistic net-proton predictions for O--O and Pb--Pb collisions at the LHC that establish a non-critical baseline for upcoming measurements.
The ideal-gas coordinate-space results are shown to be in exact agreement with the diffusion master equation of Ref.~\cite{Sakaida:2014pya}.

The paper is organized as follows.
Section~\ref{sec-loc-cons} introduces the general correlator formalism for global and local baryon conservation.
Section~\ref{sec-gaussian} specifies the Gaussian kernel ansatz and its finite-volume implementation, while Sec.~\ref{sec-baseline} derives the ideal-gas baseline and the acceptance formulas used in the phenomenological calculations.
The numerical results are presented in Sec.~\ref{sec-res}.
Technical details of the fifth- and sixth-order global correlators, the finite-volume kernels, the small-$\sigma_\eta$ limit, and the connection to diffusion dynamics are collected in the appendices.

\section{Local conservation formalism}
\label{sec-loc-cons}

\subsection{Density correlators}
\label{sec-dens-corr}

We consider fluctuations of the net baryon density $\delta\rho(\eta) \equiv \rho_B(\eta) - \mean{\rho_B(\eta)}$ in spatial rapidity $\eta$.
The $n$-point density correlator is defined as the connected (cumulant) part of the $n$-point correlation function~\cite{Speed1983}
\eq{\label{eq:Cn-def}
\mathcal{C}_n(\eta_1, \ldots, \eta_n) \equiv \left\langle \prod_{i=1}^n \delta\rho_i \right\rangle_c, \qquad n \geq 2,
}
where $\delta\rho_i \equiv \delta\rho(\eta_i)$.
The correlators are symmetric under the exchange of any two coordinates.
Integrating $\mathcal{C}_{n}$ over the full spatial rapidity range yields the $n$-th order cumulant of the net baryon number distribution,
\eq{\label{eq:Cn-kappa}
\prod_{i=1}^n \int d\eta_i \, \mathcal{C}_n(\eta_1,\ldots,\eta_n) = \kappa_n[B].
}

The connected correlators are related to the full moments through the multivariate moment-cumulant relation~\cite{Speed1983}.
Since $\mean{\delta\rho_i} = 0$, only set partitions without singleton blocks contribute, and the expressions up to fourth order read~\cite{Pratt:2019fbj}
\eq{
\mathcal{C}_2(\eta_1,\eta_2) &= \mean{\delta\rho_1 \delta\rho_2}, \label{eq:C2-mom} \\
\mathcal{C}_3(\eta_1,\eta_2,\eta_3) &= \mean{\delta\rho_1 \delta\rho_2 \delta\rho_3}, \label{eq:C3-mom} \\
\mathcal{C}_4(\eta_1,\ldots,\eta_4) &= \mean{\delta\rho_1 \delta\rho_2 \delta\rho_3 \delta\rho_4} - \mean{\delta\rho_1 \delta\rho_2}\mean{\delta\rho_3 \delta\rho_4} \nonumber \\
& - \mean{\delta\rho_1 \delta\rho_3}\mean{\delta\rho_2 \delta\rho_4} - \mean{\delta\rho_1 \delta\rho_4}\mean{\delta\rho_2 \delta\rho_3}. \label{eq:C4-mom}
}
For fifth and sixth orders, the expressions involve sums over all permutations $\sigma \in S_n$ of the index set $\{1, \ldots, n\}$:
\eq{
\mathcal{C}_5(\eta_1,\ldots,\eta_5) &= \mean{\delta\rho_1 \cdots \delta\rho_5} \nonumber \\
&- \sum_{\sigma \in S_5} \frac{\mean{\delta\rho_{\sigma_1} \delta\rho_{\sigma_2} \delta\rho_{\sigma_3}}}{3!}
\frac{\mean{\delta\rho_{\sigma_4} \delta\rho_{\sigma_5}}}{2!}, \label{eq:C5-mom}
}
\begin{widetext}
\eq{
\mathcal{C}_6(\eta_1,\ldots,\eta_6) &= \mean{\delta\rho_1 \cdots \delta\rho_6} - \sum_{\sigma \in S_6} \frac{\mean{\delta\rho_{\sigma_1} \cdots \delta\rho_{\sigma_4}}}{4!} \frac{\mean{\delta\rho_{\sigma_5} \delta\rho_{\sigma_6}}}{2!} - \frac{1}{2!}\sum_{\sigma \in S_6} \frac{\mean{\delta\rho_{\sigma_1} \delta\rho_{\sigma_2} \delta\rho_{\sigma_3}}}{3!} \frac{\mean{\delta\rho_{\sigma_4} \delta\rho_{\sigma_5} \delta\rho_{\sigma_6}}}{3!} \nonumber \\
& + \frac{2}{3!}\sum_{\sigma \in S_6} \frac{\mean{\delta\rho_{\sigma_1} \delta\rho_{\sigma_2}}}{2!} \frac{\mean{\delta\rho_{\sigma_3} \delta\rho_{\sigma_4}}}{2!} \frac{\mean{\delta\rho_{\sigma_5} \delta\rho_{\sigma_6}}}{2!}. \label{eq:C6-mom}
}
\end{widetext}
Here the factorial denominators and prefactors account for the overcounting of equivalent partitions.

\subsection{Global baryon conservation}
\label{sec-global}

In the grand-canonical ensemble, all correlations are purely local, 
\eq{
\mathcal{C}_n^{\rm gce}(\eta_1,\ldots,\eta_n) = \cc{n} \, \delta_{1\ldots n},
} where $\delta_{1\ldots n} \equiv \prod_{i=2}^n \delta(\eta_1 - \eta_i)$
and $\cc{n}$ are the baryon number susceptibilities.
In the canonical ensemble, the total baryon number is conserved exactly, requiring $\kappa_n[B_{4\pi}] = 0$ for all $n \geq 2$.
This constraint implies that the integral of $\mathcal{C}_n$ over the full volume must vanish, which is realized through non-local (balancing) contributions induced by charge conservation~\cite{Vovchenko:2024pvk}.
Each correlator $\mathcal{C}_n$ is thus decomposed into a local part proportional to $\delta_{1\ldots k}$ and non-local balancing terms that ensure the vanishing of the full-space integral.
Each term in the decomposition corresponds to a partition of the $n$ coordinate indices into groups of coinciding points, with the coefficient determined by the corresponding joint cumulant evaluated in the thermodynamic limit.
The canonical density correlators for uniform systems with global baryon conservation have been derived in Ref.~\cite{Vovchenko:2024pvk} up to the fourth order.
Here we present the generalization to higher orders.

We use the shorthand $\delta_{i,j} \equiv \delta(\eta_i - \eta_j)$ and $\delta_{i,j,k} \equiv \delta(\eta_i - \eta_j)\delta(\eta_i - \eta_k)$, and similarly for higher-order products.
The second- and third-order correlators read~\cite{Vovchenko:2024pvk}
\eq{
\mathcal{C}_2(\eta_1,\eta_2) &= \cc{2} \, \delta_{1,2} - \frac{\cc{2}}{V}, \label{eq:C2-global} \\
\mathcal{C}_3(\eta_1,\eta_2,\eta_3) &= \cc{3} \, \delta_{1,2,3} - \frac{\cc{3}}{V}\left[\delta_{1,2} + \delta_{1,3} + \delta_{2,3}\right] + \frac{2\cc{3}}{V^2}. \label{eq:C3-global}
}
The fourth-order correlator is~\cite{Vovchenko:2024pvk}
\begin{widetext}
\eq{
\mathcal{C}_4(\eta_1,\ldots,\eta_4) &= \cc{4} \, \delta_{1,2,3,4} - \frac{\cc{4}}{V}\left[\delta_{1,2,3} + \delta_{1,2,4} + \delta_{1,3,4} + \delta_{2,3,4}\right] - \frac{(\cc{3})^2}{\cc{2} V}\left[\delta_{1,2}\delta_{3,4} + \delta_{1,3}\delta_{2,4} + \delta_{1,4}\delta_{2,3}\right] \nonumber \\
&+ \frac{1}{V^2}\left[\cc{4} + \frac{(\cc{3})^2}{\cc{2}}\right]\left[\delta_{1,2} + \delta_{1,3} + \delta_{1,4} + \delta_{2,3} + \delta_{2,4} + \delta_{3,4}\right] - \frac{3}{V^3}\left[\cc{4} + \frac{(\cc{3})^2}{\cc{2}}\right]. \label{eq:C4-global}
}
The fifth- and sixth-order correlators $\mathcal{C}_5$ and $\mathcal{C}_6$, derived in this work, are given in Appendix~\ref{app:C56-local}.
\end{widetext}
The structure of each correlator reflects the hierarchy of set partitions: the leading term is the fully local contribution proportional to $\delta_{1\ldots n}$, followed by increasingly non-local balancing terms with higher powers of $1/V$, down to the fully non-local constant. Together, these terms ensure the vanishing of the full-volume integral.

\subsection{Local baryon conservation}
\label{sec-loc-cons-general}

In the global conservation case, the non-local balancing terms in $\mathcal{C}_n$ are spatially uniform, implying that baryon charges at opposite ends of the fireball are correlated as strongly as those nearby.
This is unphysical when the fireball extends over a rapidity range larger than the causal correlation length~\cite{Castorina:2013mba}.
Local conservation remedies this by confining the balancing correlations to a finite range in rapidity. 
This is encoded in local-conservation kernels, which we denote by $\varkappa_n(\eta_1,\ldots,\eta_n)$, following Ref.~\cite{Vovchenko:2024pvk}, reserving $\kappa_n$ for cumulants.

Consider first the two-point correlator.
As shown in Ref.~\cite{Vovchenko:2024pvk}, local conservation is introduced by replacing the constant balancing term $-\cc{2}/V$ in Eq.~\eqref{eq:C2-global} with a spatially dependent kernel,
\eq{\label{eq:C2-local}
\mathcal{C}_2(\eta_1,\eta_2) = \cc{2}\left[\delta_{1,2} - \frac{\varkappa_2(\eta_1,\eta_2)}{V}\right].
}
Here $\varkappa_2(\eta_1,\eta_2)$ is a symmetric, positive-definite balancing function that describes how a baryon charge at $\eta_1$ is compensated in the vicinity of $\eta_2$.
It satisfies the sum rule $\int d\eta_2\, \varkappa_2(\eta_1,\eta_2) = V$, which ensures the conservation constraint $\int d\eta_2\, \mathcal{C}_2 = 0$.
In the global conservation limit $\varkappa_2 \to 1$, Eq.~\eqref{eq:C2-global} is recovered.
In Ref.~\cite{Vovchenko:2024pvk}, $\varkappa_2$ was modeled as a Gaussian in spatial rapidity with width $\sigma_\eta$; we defer the specific functional form used here to Sec.~\ref{sec-gaussian}.

The extension to the three-point correlator introduces a three-point kernel $\varkappa_3(\eta_1,\eta_2,\eta_3)$.
The local conservation analog of Eq.~\eqref{eq:C3-global} reads
\eq{\label{eq:C3-local}
\mathcal{C}_3(\eta_1,\eta_2,\eta_3) &= \cc{3} \, \delta_{1,2,3}  - \frac{\cc{3}}{V}\Big[\delta_{1,2}\,\varkappa_2(\eta_1,\eta_3) \nonumber \\
&
 \qquad + \delta_{1,3}\,\varkappa_2(\eta_1,\eta_2)  + \delta_{2,3}\,\varkappa_2(\eta_2,\eta_1)\Big] \nonumber \\
& \qquad + \frac{2\cc{3}}{V^2}\,\varkappa_3(\eta_1,\eta_2,\eta_3) .
}
The structure mirrors the global case: each power of $1/V$ appearing in the balancing terms of Eq.~\eqref{eq:C3-global} is replaced by the corresponding kernel, $1/V^k \to \varkappa_{k+1}/V^k$.
The kernel $\varkappa_3$ is not independent of $\varkappa_2$: its normalization is fixed by the conservation constraint $\int d\eta_3\, \mathcal{C}_3 = 0$.
Evaluating this integral and using $\int d\eta_3\, \varkappa_2(\eta_i,\eta_3) = V$, one obtains the \emph{reduction condition}
\eq{\label{eq:reduction}
\int d\eta_3\, \varkappa_3(\eta_1,\eta_2,\eta_3) = V\,\varkappa_2(\eta_1,\eta_2).
}
Thus, integrating out one coordinate of $\varkappa_3$ recovers $\varkappa_2$, up to a factor of $V$.
The same pattern continues to higher orders: for each correlator $\mathcal{C}_n$, the balancing terms with $k$ non-coinciding groups involve the kernel $\varkappa_k$, and the conservation constraint $$\int d\eta_n\, \mathcal{C}_n = 0$$ yields the general reduction condition
\eq{\label{eq:reduction-general}
\int d\eta_n\, \varkappa_n(\eta_1,\ldots,\eta_n) = V\,\varkappa_{n-1}(\eta_1,\ldots,\eta_{n-1}).
}
While the reduction conditions~\eqref{eq:reduction-general} fix the marginals of $\varkappa_n$ in terms of $\varkappa_{n-1}$, they do not uniquely determine the full $n$-point kernels for $n \geq 3$.
We close the hierarchy by adopting a symmetric Gaussian-cluster ansatz introduced in Sec.~\ref{sec-npoint-kernels}.
The explicit expressions for $\mathcal{C}_4$ through $\mathcal{C}_6$ with general kernels $\varkappa_n$ follow the same substitution rule $1/V^k \to \varkappa_{k+1}/V^k$ applied to the global conservation correlators in Eq.~\eqref{eq:C4-global}; the global limit is recovered by setting $\varkappa_n \to 1$.
In each correlator, the terms are organized by the set partitions of the $n$ coordinate indices: the leading fully local term $\propto \delta_{1\ldots n}$ is followed by balancing terms of increasing non-locality, with each power $1/V^k$ in the global formula promoted to a kernel factor $\varkappa_{k+1}/V^k$ in the local one.
\begin{widetext}
The fourth-order correlator reads
\eq{
&\mathcal{C}_4(\eta_1,\dots,\eta_4)  = \cc{4} \delta_{1,2,3,4} - \frac{\cc{4}}{3!V}\sum\limits_{\sigma \in S_4} \delta_{\sigma_1,\sigma_2,\sigma_3}\varkappa_2(\eta_{\sigma_1},\eta_{\sigma_4}) - \frac{(\cc{3})^2}{(2!)^2 \cdot 2!\,\cc{2} V} \sum\limits_{\sigma \in S_4}\delta_{\sigma_1,\sigma_2} \delta_{\sigma_3,\sigma_4}\varkappa_2(\eta_{\sigma_1},\eta_{\sigma_3}) \nonumber \\&+ \frac{1}{2!\cdot 2!\,V^2} \left[\cc{4} + \frac{(\cc{3})^2}{\cc{2}} \right] \sum\limits_{\sigma \in S_4} \delta_{\sigma_1,\sigma_2}\varkappa_3(\eta_{\sigma_1},\eta_{\sigma_3},\eta_{\sigma_4}) - \frac{3}{V^3} \left[\cc{4} + \frac{(\cc{3})^2}{\cc{2}} \right] \varkappa_4(\eta_1,\eta_2,\eta_3,\eta_4).
}
The explicit expressions for $\mathcal{C}_5$ and $\mathcal{C}_6$ are given in Appendix~\ref{app:C56-local}.
 \end{widetext}

\section{Gaussian local conservation}
\label{sec-gaussian-loc}

\subsection{Two-point kernel}
\label{sec-gaussian}

The formalism developed above holds for any choice of $\varkappa_2$ satisfying the appropriate symmetry and normalization properties.
Following Ref.~\cite{Vovchenko:2024pvk}, we model the two-point balancing kernel as a Gaussian in spatial rapidity,
\eq{\label{eq:kappa2-inf}
\varkappa_2(\eta_1,\eta_2) \propto \exp\!\left[-\frac{(\eta_1-\eta_2)^2}{2\sigma_\eta^2}\right],
}
where $\sigma_\eta$ characterizes the range of local baryon conservation.
In the limit $\sigma_\eta \to \infty$ the kernel becomes flat and the global conservation result is recovered.

In a finite system of size $V = 2\eta_{\rm max}$, the kernel must respect the boundary conditions of the fireball.
In Ref.~\cite{Vovchenko:2024pvk}, a minimum-image convention was adopted: the distance $|\eta_1 - \eta_2|$ in Eq.~\eqref{eq:kappa2-inf} is replaced by the shortest distance on a periodic interval of length $V$.
This prescription is adequate when $\sigma_\eta \ll \eta_{\rm max}$, but has two limitations.
First, it treats the fireball boundaries as periodic, which is unphysical: baryon charge reaching one end of the fireball should not reappear at the other.
Second, the minimum-image convention does not generalize straightforwardly to $n$-body kernels involving multiple pairwise distances.

To overcome these limitations, we consider two alternative boundary conditions that are implemented through exact summation over all images of the Gaussian, using the Poisson summation formula (see Appendix~\ref{jacobi_AP} for details).

The first option is \emph{full periodic} boundary conditions, in which the Gaussian is periodically continued across all images.
This yields a closed-form expression in terms of the Jacobi theta function $\theta_3$~[Eq.~\eqref{eq:pbc_kernel}].
While systematic and naturally generalizable to higher-order kernels, fully periodic boundary conditions introduce the same unphysical artifact as the minimum-image convention: when the kernel is centered near one edge of the fireball, a spurious ``echo'' of the correlation appears at the opposite boundary due to the periodic images wrapping around the system.

The second option is \emph{reflecting} boundary conditions, implemented by doubling the periodicity interval and including mirror images at the boundaries (see Appendix~\ref{jacobi_AP}).
The resulting kernel, expressed through Eq.~\eqref{eq:apbc_kernel}, is strongly suppressed at the boundary opposite to the source, which is the physically expected behavior for a fireball with hard walls: charge correlations are strongest near the source and decay toward zero at the far end.

\begin{figure*}
    \centering
    \includegraphics[width=0.83\linewidth]{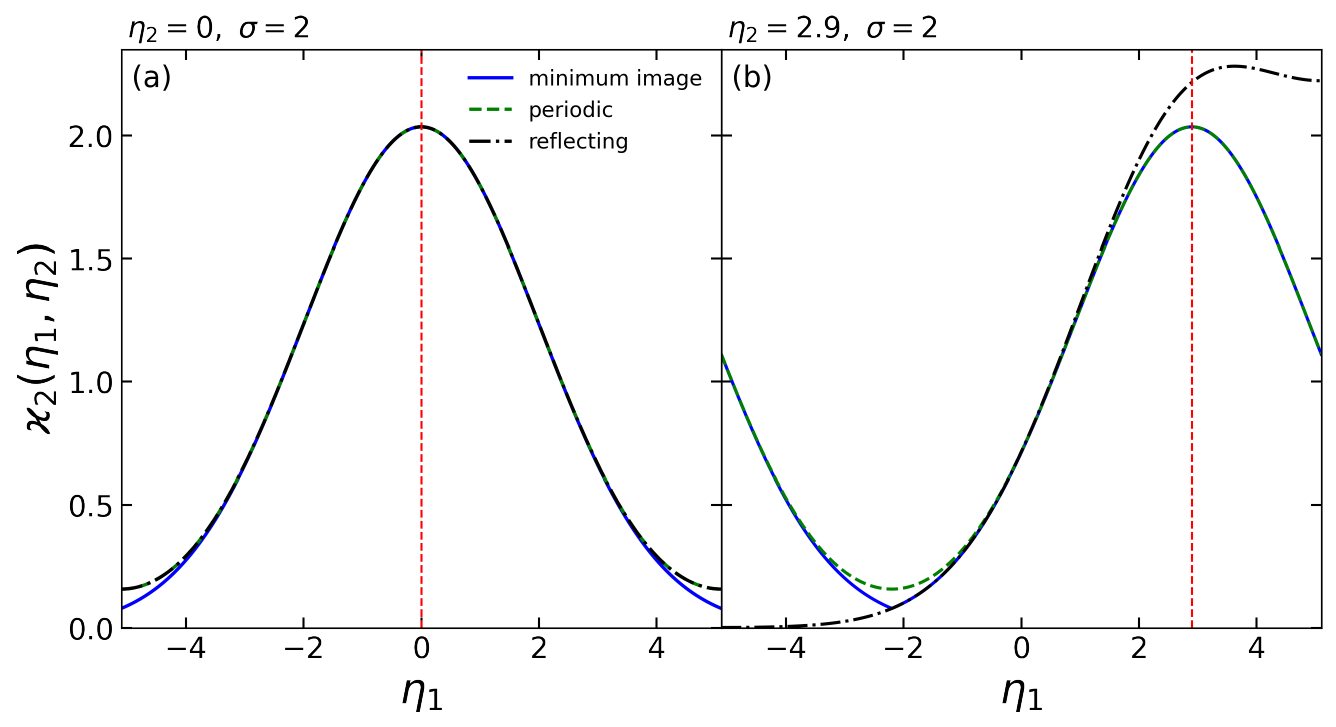}
    \caption{The two-point conservation kernel $\varkappa_2(\eta_1,\eta_2)$ as a function of $\eta_1$ at fixed $\eta_2$ for $\sigma_\eta = 2$ and $\eta_{\rm max} = 5.1$.
    Three boundary conditions are compared: minimum-image periodic (solid blue), full periodic (dashed green), and reflecting (dash-dotted black).
    The vertical dashed red line indicates the position of $\eta_2$.
    (a)~$\eta_2 = 0$ (centered): all three prescriptions coincide.
    (b)~$\eta_2 = 2.9$ (shifted toward the boundary): the periodic kernel develops a spurious enhancement at the far boundary due to wrapping, while the reflecting kernel decays toward zero there.}
    \label{fig:BC}
\end{figure*}

Figure~\ref{fig:BC} compares the three prescriptions for the two-point kernel $\varkappa_2(\eta_1,\eta_2)$ as a function of $\eta_1$ at fixed $\eta_2$, with $\sigma_\eta = 2$ and $\eta_{\rm max} \simeq 5.1$.
When $\eta_2$ is at the center of the fireball~[panel~(a)], all three approaches agree, as the kernel tails are sufficiently suppressed before reaching the boundaries.
When $\eta_2$ is shifted to $\eta_2 = 2.9$, close to the boundary~[panel~(b)], the differences become pronounced: the minimum-image and fully periodic kernels (solid blue and dashed green) both drop toward zero near the boundary closest to the source, missing the contribution from mirror images; additionally, the periodic kernel develops a spurious enhancement at the far boundary due to the same wrapping artifact.
In contrast, the reflecting kernel (dash-dotted black) is enhanced near the close boundary by the mirror image of the source and decays smoothly toward zero at the far boundary, as expected physically for a fireball with hard walls.
Since the reflecting kernel provides the most physical description of the fireball boundaries, we adopt it as the default choice in this work.
We note that for observables evaluated at midrapidity, where the kernel is centered far from the boundaries, all three prescriptions yield equivalent results and this is why the analysis of second-order cumulants in Ref.~\cite{Vovchenko:2024pvk} was insensitive to the choice of boundary conditions.
The distinction becomes relevant, however, for higher-order cumulants, which involve $n$-point kernels $\varkappa_n$ with multiple coordinates that can probe different regions of the fireball simultaneously.

\subsection{\texorpdfstring{$n$-point kernels}{n-point kernels}}
\label{sec-npoint-kernels}

To evaluate the local conservation correlators $\mathcal{C}_n$ for $n \geq 3$, one requires the higher-order kernels $\varkappa_n(\eta_1,\ldots,\eta_n)$ to satisfy the reduction conditions~\eqref{eq:reduction-general}.
We adopt a symmetric Gaussian cluster ansatz, in which $n$ coordinates are mutually correlated through a Gaussian weight depending on all pairwise distances:
\eq{\label{eq:kappa_n}
\varkappa_n(\eta_1, \ldots, \eta_n) \propto A_n \exp\!\left[ -\frac{1}{n \sigma_{\eta}^2} \sum_{1 \le i < j \le n} (\eta_i - \eta_j)^2 \right].
}
This ansatz is the natural $n$-body generalization of the Gaussian two-point kernel~\eqref{eq:kappa2-inf}: for $n=2$ it reduces to $\varkappa_2 \propto \exp[{-(\eta_1-\eta_2)^2/(2\sigma_\eta^2)}]$, and the width $\sigma_\eta$ is shared across all orders, so that the single parameter extracted from second-order cumulants determines the entire kernel hierarchy.

In infinite volume, integrating out any one coordinate in Eq.~\eqref{eq:kappa_n} yields a Gaussian of the same form in the remaining $n-1$ coordinates, so the reduction condition~\eqref{eq:reduction-general} is automatically satisfied with
\eq{\label{eq:An}
A_n = \frac{1}{\sqrt{n}} \left( \frac{V}{\sqrt{\pi}\,\sigma_{\eta}} \right)^{n-1}.
}

In a finite system, the boundary conditions are implemented by summing over images of the Gaussian, as for $\varkappa_2$.
For periodic boundary conditions, the $n$-point kernel reads
\begin{widetext}
\eq{\label{eq:kappa_n_pbc}
\begin{split}
\varkappa^{\text{PBC}}_{n}(\eta_1, \dots, \eta_n)
&= A_n \!\!\sum_{k_2, \dots, k_n = -\infty}^{\infty}
\exp\!\left[
- \frac{1}{n\sigma_\eta^2}
\sum_{1 \le i < j \le n} (\tilde{\eta}_i - \tilde{\eta}_j)^2
\right]\!,
\end{split}
}
\end{widetext}
with shifted coordinates $\tilde{\eta}_j = \eta_j + 2k_j \eta_{\rm max}$ for $j \geq 2$ and $\tilde{\eta}_1 = \eta_1$.
For reflecting boundary conditions, each image coordinate carries an additional reflection index $s_j \in \{0,1\}$, with $\tilde{\eta}_j = \eta_j + 4k_j\eta_{\rm max}$ (direct, $s_j=0$) or $\tilde{\eta}_j = 2\eta_{\rm max} - \eta_j + 4k_j\eta_{\rm max}$ (reflected, $s_j=1$), and the kernel reads: 
\begin{widetext}
\eq{\label{eq:kappa_n_rbc}
\begin{split}
\varkappa^{\text{RBC}}_{n}(\eta_1, \dots, \eta_n)
&= A_n \!\!\sum_{k_2, \dots, k_n = -\infty}^{\infty} \sum_{s_2,\dots,s_n = 0}^1
\exp\!\left[
- \frac{1}{n\sigma_\eta^2}
\sum_{1 \le i < j \le n} (\tilde{\eta}_i - \tilde{\eta}_j)^2
\right]\!,
\end{split}
}
\end{widetext}
See Appendix~\ref{jacobi_AP} for details and a compact analytical expression for the 2-point kernel in terms of Jacobi theta functions.
Unlike the two-point case, the $n$-point kernels with finite-system boundary conditions do not admit a simple analytical form for $n \geq 3$.
In practice, the image sums are truncated at a finite number of terms; convergence is controlled by the ratio $\sigma_\eta / \eta_{\rm max}$, with fewer images needed when the kernel is narrow relative to the system size.
We also evaluate the acceptance integrals $\mathcal{J}_n$ using the common-source representation (see Appendix~\ref{app:small-sigma}), which reduces the $n$-dimensional integral to a one-dimensional quadrature over the source coordinate.
For the parameter range used in this work ($\sigma_\eta \lesssim 2$, $\eta_{\rm max} = 5.1$), truncating the one-body image sum at $\sim 10$ terms is fully sufficient for convergence.

\section{Ideal gas baseline and kinematic cuts}
\label{sec-baseline}

\subsection{Poisson baseline}

In a non-interacting hadron resonance gas, the baryon number susceptibilities follow the Skellam baseline (Poisson statistics).
Baryons and antibaryons are produced independently with mean numbers $\mean{B}$ and $\mean{\bar{B}}$, yielding
\eq{
\cc{2k} = \frac{\mean{B{+}\bar{B}}}{V}, \qquad \cc{2k+1} = \frac{\mean{B-\bar{B}}}{V}, \label{eq:skellam}
}
for all positive integers $k$.
That is, even-order susceptibilities measure the total baryon-plus-antibaryon density, while odd-order susceptibilities measure the net-baryon density.
At the LHC, the net-baryon density is negligibly small ($\mu_B \approx 0$), so that $\cc{2k+1} \approx 0$ and all even susceptibilities coincide, $\cc{2} = \cc{4} = \cc{6} = \cdots \equiv \mean{B{+}\bar{B}}/V$.

The ideal gas provides a natural baseline for two reasons.
First, it establishes the non-critical reference against which deviations due to the chiral crossover or other medium effects can be identified.
Second, the absence of local many-body correlations in the ideal gas, where all correlations are either self-correlations at coinciding coordinates or non-local correlations induced by conservation laws, greatly simplifies the modeling of kinematic acceptance effects.
In particular, the local terms in the density correlators $\mathcal{C}_n$ are purely single-particle self-correlations (proportional to $\delta_{1\ldots k}$), so each self-correlation term contributes only a single power of the acceptance probability $p(\eta)$ regardless of the order $k$ of the coinciding delta function~\cite{Vovchenko:2024pvk}.
In the presence of local many-body correlations from the equation of state, the local terms would involve genuine multi-particle correlations at coinciding coordinates, requiring higher powers of $p(\eta)$ weighted by distinct susceptibility combinations. This complication does not arise in the ideal gas we study here.

\subsection{\texorpdfstring{Simplified correlators at $\mu_B = 0$}{Simplified correlators at muB = 0}}

Setting the odd susceptibilities to zero and using $\cc{2} = \cc{4} = \cc{6} = \mean{B{+}\bar{B}}/V$, the density correlators simplify considerably.
The odd-order correlators vanish identically, $\mathcal{C}_3 = \mathcal{C}_5 = 0$, and all partition terms whose coefficients depend solely on odd susceptibilities drop out, in particular the $(2,2)$ partition in $\mathcal{C}_4$ and the $(4,2)$ and $(2,2,2)$ partitions in $\mathcal{C}_6$.
For the second- and fourth-order correlators with local conservation one obtains
\eq{
\mathcal{C}_2(\eta_1,\eta_2) &= \frac{\mean{B{+}\bar{B}}}{V}\left[\delta_{1,2} - \frac{\varkappa_2(\eta_1,\eta_2)}{V}\right], \label{eq:C2-idg} \\
\mathcal{C}_4(\eta_1,\ldots,\eta_4) &= \frac{\mean{B{+}\bar{B}}}{V}\bigg[\delta_{1,2,3,4} \nonumber \\
&- \frac{1}{V}\Big(\delta_{1,2,3}\,\varkappa_2^{1,4} + \delta_{1,2,4}\,\varkappa_2^{1,3} \nonumber \\
& \qquad + \delta_{1,3,4}\,\varkappa_2^{1,2} + \delta_{2,3,4}\,\varkappa_2^{2,1}\Big) \nonumber \\
&+ \frac{1}{V^2}\Big(\delta_{1,2}\,\varkappa_3^{1,3,4} + \delta_{1,3}\,\varkappa_3^{1,2,4} + \delta_{1,4}\,\varkappa_3^{1,2,3} \nonumber \\
& \qquad + \delta_{2,3}\,\varkappa_3^{2,1,4} + \delta_{2,4}\,\varkappa_3^{2,1,3} + \delta_{3,4}\,\varkappa_3^{3,1,2}\Big) \nonumber \\
&- \frac{3}{V^3}\,\varkappa_4(\eta_1,\eta_2,\eta_3,\eta_4)\bigg], \label{eq:C4-idg}
}
where $\varkappa_k^{i,j,\ldots} \equiv \varkappa_k(\eta_i,\eta_j,\ldots)$ is a shorthand for the kernel arguments.
In each case, because all even susceptibilities coincide in the ideal gas, the entire correlator factors into $\mean{B{+}\bar{B}}/V$ times a combination of conservation kernels and delta functions that depends only on the spatial coordinates and the kernel width~$\sigma_\eta$.

\subsection{Acceptance}

To compute observable cumulants $\kappa_n[B]_{\rm acc}$ inside a limited acceptance, we integrate the spatial correlators weighted by a single-particle acceptance function $p(\eta)$ at each coordinate.
The function $p(\eta)$ encodes the probability that a baryon at spatial rapidity $\eta$ is detected.
In the simplest case, $p(\eta)$ models a coordinate-space rapidity cut as a step function, $p(\eta) = \Theta(\eta_{\rm cut} - |\eta|)$, selecting baryons within $|\eta| < \eta_{\rm cut}$.
More generally, $p(\eta)$ can incorporate kinematic cuts and detector efficiencies, including the fact that experiments typically measure protons rather than all baryons.
In that case, $p(\eta)$ is the probability that a baryon at rapidity $\eta$ is a proton (or antiproton) that falls within the experimental momentum acceptance and is reconstructed with the appropriate efficiency~\cite{Vovchenko:2024pvk}.
In the ideal gas, the binomial factorization discussed above allows us to define the acceptance-weighted integrals of the $n$-body kernels,
\eq{\label{eq:Jn-def}
\mathcal{J}_n \equiv \frac{1}{V^{n-1}} \int d\eta_1 \ldots d\eta_n \, p(\eta_1) \ldots p(\eta_n) \, \varkappa_n(\eta_1, \ldots, \eta_n),
}
with $\mathcal{J}_1 = \int p(\eta)\, d\eta = V \mean{p}$.
For the Gaussian-cluster ansatz, the integrals $\mathcal{J}_n$ admit an efficient common-source representation that reduces the $n$-dimensional integral above to a one-dimensional quadrature in an auxiliary source coordinate; see Appendix~\ref{app:small-sigma}.
At the LHC ($\mu_B \approx 0$), the odd-order cumulants vanish identically, $\kappa_1 = \kappa_3 = \kappa_5 = 0$, while the even-order cumulants factor into $\cc{2} = \mean{B{+}\bar{B}}/V$ (the mean baryon-plus-antibaryon density in full phase space) times a purely geometric polynomial in $\mathcal{J}_n$.
The Skellam (Poisson) baseline for the net-baryon variance at $\mu_B = 0$ is equal to the mean accepted baryon-plus-antibaryon multiplicity,
\eq{\label{eq:Npm-acc}
\Nacc \equiv \mean{B{+}\bar{B}}_{\rm acc} = \frac{\mean{B{+}\bar{B}}}{V}\,\mathcal{J}_1.
}
The even-order net-baryon cumulants read
\eq{
\kappa_2 &= \frac{\mean{B{+}\bar{B}}}{V} \left[ \mathcal{J}_1 - \mathcal{J}_2 \right], \label{eq:k2-idg} \\
\kappa_4 &= \frac{\mean{B{+}\bar{B}}}{V} \left[ \mathcal{J}_1 - 4\, \mathcal{J}_2 + 6\, \mathcal{J}_3 - 3\, \mathcal{J}_4 \right], \label{eq:k4-idg} \\
\kappa_6 &= \frac{\mean{B{+}\bar{B}}}{V} \big[ \mathcal{J}_1 - 16\, \mathcal{J}_2 + 75\, \mathcal{J}_3 \nonumber \\
& \quad - 150\, \mathcal{J}_4 + 135\, \mathcal{J}_5 - 45\, \mathcal{J}_6 \big]. \label{eq:k6-idg}
}
Since all cumulants are proportional to the same susceptibility, the commonly measured ratios simplify:
\eq{
\frac{\kappa_2}{\Nacc} &= 1 - \frac{\mathcal{J}_2}{\mathcal{J}_1}\,, \label{eq:k2-ratio} \\[6pt]
\frac{\kappa_4}{\kappa_2} &= \frac{\mathcal{J}_1 - 4\,\mathcal{J}_2 + 6\,\mathcal{J}_3 - 3\,\mathcal{J}_4}{\mathcal{J}_1 - \mathcal{J}_2}\,, \label{eq:k4k2-ratio} \\[6pt]
\frac{\kappa_6}{\kappa_2} &= \frac{1}{\mathcal{J}_1 - \mathcal{J}_2}\big[\mathcal{J}_1 - 16\,\mathcal{J}_2 + 75\,\mathcal{J}_3 \nonumber \\
& \quad - 150\,\mathcal{J}_4 + 135\,\mathcal{J}_5 - 45\,\mathcal{J}_6\big]. \label{eq:k6k2-ratio}
}
These ratios depend only on the acceptance function $p(\eta)$ and the conservation kernel width $\sigma_\eta$, and are independent of the overall baryon density.
The ratio $\kappa_2/\Nacc$ quantifies the suppression of the variance relative to the Skellam baseline of unity, while $\kappa_4/\kappa_2$ and $\kappa_6/\kappa_2$ measure the non-Gaussianity of the net-baryon distribution induced by local conservation.
In the global conservation limit, $\mathcal{J}_n \to \mathcal{J}_1^n / V^{n-1}$, and these expressions reduce to the known results~\cite{Vovchenko:2024pvk}.

\subsection{Baryon and antibaryon cumulants}

The formulas above give cumulants of the net-baryon number $B{-}\bar{B}$.
Experiments also measure proton-only (baryon-only) fluctuations and baryon--antibaryon correlations.
At $\mu_B = 0$ and in the ideal gas, the density correlators can be decomposed into species components.
Each baryon carries baryon number $+1$ and each antibaryon $-1$, so the fraction of the balancing charge carried by like-type particles is $1/2$.
The second-order baryon--baryon and baryon--antibaryon density correlators read
\eq{
\mathcal{C}_2^{BB}(\eta_1,\eta_2) &= \frac{\mean{B}}{V}\left[\delta_{1,2} - \frac{\varkappa_2(\eta_1,\eta_2)}{2V}\right], \label{eq:C2-BB} \\
\mathcal{C}_{11}^{B\bar{B}}(\eta_1,\eta_2) &= \frac{\mean{B{+}\bar{B}}}{4V^2}\,\varkappa_2(\eta_1,\eta_2), \label{eq:C11-BBbar}
}
where $\mean{B} = \mean{\bar{B}} = \mean{B{+}\bar{B}}/2$ at $\mu_B = 0$.
The baryon--baryon correlator~\eqref{eq:C2-BB} has a balancing term reduced by the factor $1/2$ relative to the net-baryon case~\eqref{eq:C2-idg}, reflecting the fact that only half of the balancing charge (the antibaryon partner) anticorrelates with the observed baryon.
The cross-correlator~\eqref{eq:C11-BBbar} is purely non-local and positive: conservation induces correlations between unlike baryons by confining each baryon--antibaryon pair within a distance $\sim\sigma_\eta$.
The total baryon-plus-antibaryon correlator $\mathcal{C}_2^{B+\bar{B}} = 2\mathcal{C}_2^{BB} + 2\mathcal{C}_{11}^{B\bar{B}} = (\mean{B{+}\bar{B}}/V)\,\delta_{1,2}$ is purely Poissonian: the conservation-induced anticorrelation among like baryons is exactly canceled by the positive cross-correlation between baryons and antibaryons.
This cancellation holds in any acceptance within the thermodynamic limit used here, giving $\kappa_2[B{+}\bar{B}] = \Nacc$, and reflects the fact that baryon number conservation constrains the difference $B{-}\bar{B}$ but leaves the total $B{+}\bar{B}$ unaffected.
We note that in a finite volume (finite multiplicity) the exact canonical ensemble gives a super-Poissonian total variance, $\kappa_2[B{+}\bar{B}]/\Nacc > 1$~\cite{Begun:2004gs}, with corrections of order $1/\mean{B{+}\bar{B}}$ that are negligible for the systems considered here.
Thus, measuring both $\kappa_2[B{-}\bar{B}]$ and $\kappa_2[B{+}\bar{B}]$ provides a direct test of the conservation mechanism: any deviation of $\kappa_2[B{+}\bar{B}]$ from Poisson beyond the expected finite-multiplicity correction would signal contributions beyond the ideal gas, such as baryon-antibaryon annihilation~\cite{Savchuk:2021aog} or critical fluctuations.

Integrating with the acceptance function $p(\eta)$ gives
\eq{
\kappa_2[B] &= \frac{\mean{B}}{V}\left[\mathcal{J}_1 - \frac{\mathcal{J}_2}{2}\right], \label{eq:k2-B} \\
\mathrm{Cov}(B,\bar{B})_{\rm acc} &= \frac{\mean{B}}{V}\,\frac{\mathcal{J}_2}{2}\,. \label{eq:cov-BBbar}
}
The baryon-only scaled variance reads
\eq{ \label{eq:k2B-ratio}
\frac{\kappa_2[B]}{\mean{B}_{\rm acc}} = 1 - \frac{\mathcal{J}_2}{2\mathcal{J}_1} = \frac{1}{2}\left(1 + \frac{\kappa_2[B{-}\bar{B}]}{\Nacc}\right)\!,
}
which is always closer to unity (less suppressed) than the net-baryon ratio $\kappa_2[B{-}\bar{B}]/\Nacc$.
These expressions reproduce the known results for particle number fluctuations in the canonical ensemble~\cite{Begun:2004gs}.

\section{Results}
\label{sec-res}

\subsection{Coordinate-space cumulants}

We first present results for coordinate-space cumulant ratios, obtained using a step-function acceptance $p(\eta) = \Theta(\eta_{\rm cut} - |\eta|)$, as a function of the acceptance fraction $\alpha = \eta_{\rm cut}/\eta_{\rm max}$.
This setup isolates the effect of local conservation from kinematic smearing and allows direct comparison with the diffusion model of Ref.~\cite{Sakaida:2014pya}.
All results in this subsection correspond to the ideal gas at $\mu_B = 0$ ($\cc{3} = 0$) with reflecting boundary conditions and $\eta_{\rm max} = 5.1$~\cite{Parra:2025fse}, representative of $\sNN = 5.02$~TeV Pb--Pb collisions.

\begin{figure*}
    \centering
    \includegraphics[width=0.98\linewidth]{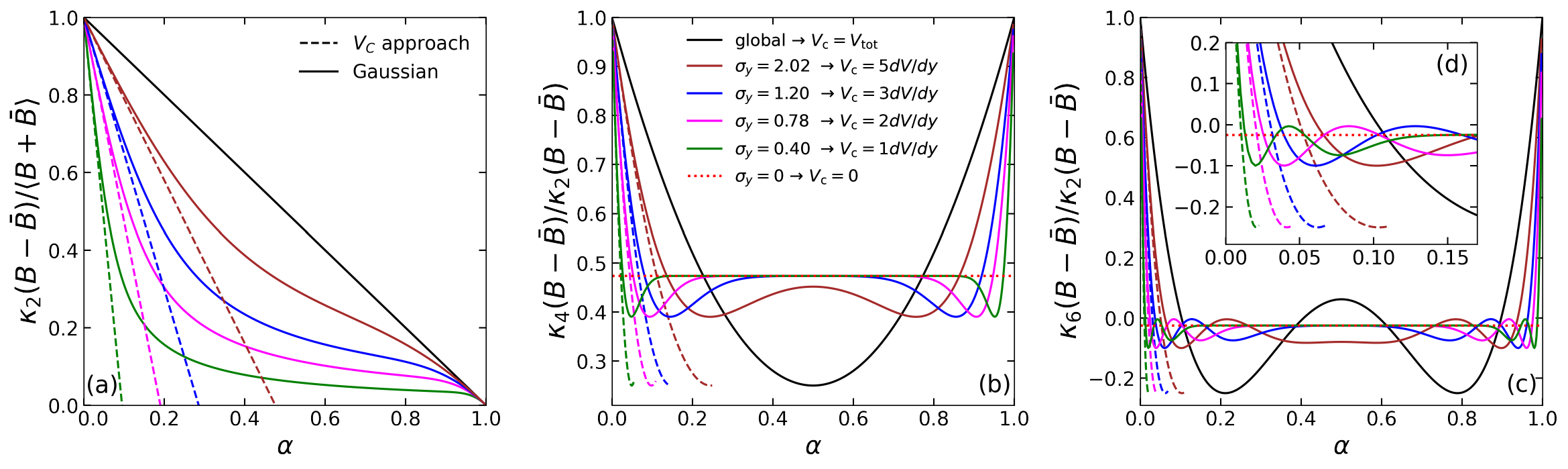}
    \caption{Coordinate-space cumulant ratios $\kappa_2/\Nacc$ (left), $\kappa_4/\kappa_2$ (middle), and $\kappa_6/\kappa_2$ (right) as a function of the acceptance fraction $\alpha = \eta_{\rm cut}/\eta_{\rm max}$ for the ideal gas at $\mu_B = 0$, where $\Nacc \equiv \mean{B{+}\bar{B}}_{\rm acc}$.
    Solid lines correspond to the Gaussian local-conservation kernel with widths $\sigma_\eta = 2.02, 1.20, 0.78, 0.40$, and the global conservation limit (solid black).
    Dashed lines show the equivalent $V_c$ approach of Ref.~\cite{Vovchenko:2024pvk} with $V_c = k\,dV/dy$, $k = 5, 3, 2, 1$.
    Reflecting boundary conditions are used throughout.
    }
    \label{fig:k2k4-coord}
\end{figure*}

Figure~\ref{fig:k2k4-coord} shows the ratios $\kappa_2/\Nacc$, $\kappa_4/\kappa_2$, and $\kappa_6/\kappa_2$ for several values of the conservation width $\sigma_\eta$.
In the left panel, $\kappa_2/\Nacc$ starts at unity (Skellam baseline) for vanishing acceptance and decreases monotonically toward zero at $\alpha = 1$, where the full-space conservation constraint $\kappa_2 = 0$ is enforced.
For global conservation ($\sigma_\eta \to \infty$), the suppression is linear, $\kappa_2/\Nacc = 1 - \alpha$, as is well known~\cite{Vovchenko:2024pvk}.
As $\sigma_\eta$ decreases, the suppression strengthens at intermediate $\alpha$: a narrow conservation kernel confines the balancing antibaryon close to the original baryon and simultaneously depletes nearby same-sign baryons, so that even a small acceptance window captures a larger fraction of both effects, reducing the net-baryon variance.

The middle panel shows $\kappa_4/\kappa_2$, which exhibits a richer structure.
The ratio approaches unity at both $\alpha \to 0$ (Poisson limit) and $\alpha \to 1$, where all curves converge regardless of $\sigma_\eta$, since the full-space cumulants are determined by the global conservation constraint alone.
For large $\sigma_\eta$ (close to global conservation), a single minimum develops near $\alpha = 0.5$.
As $\sigma_\eta$ decreases, the minimum splits into a characteristic double-minimum structure with a plateau region around $\alpha \sim 0.5$ where $\kappa_4/\kappa_2$ is approximately constant.
The plateau value can be understood analytically in the limit $\sigma_\eta \ll \eta_{\rm cut}$ and $\sigma_\eta \ll \eta_{\rm max} - \eta_{\rm cut}$.
In this regime, both $\kappa_2$ and $\kappa_4$ are dominated by boundary-layer effects: only baryons emitted within $O(\sigma_\eta)$ of the acceptance edges $\pm\eta_{\rm cut}$ have their balancing charge leak outside.
As shown in Appendix~\ref{app:small-sigma}, the Gaussian cluster kernel admits a common-source representation in which the $n$ coordinates of $\varkappa_n$ are generated as independent Gaussian offsets from a shared source position $X$.
The acceptance integrals then take the form $\mathcal{J}_n = 2\eta_{\rm cut} - \lambda_n\,\sigma_\eta + O(\sigma_\eta^2)$.
The leading correction $\lambda_n\,\sigma_\eta$ measures the average rapidity range spanned by $n$ correlated coordinates sharing a common source: $\lambda_n = \sqrt{2}\,\mathbb{E}[\max(Z_1,\ldots,Z_n)]$, where the $Z_i$ are independent standard normal variables.
Physically, a larger $\lambda_n$ means the $n$-body kernel extends further from the source, increasing the probability that at least one coordinate falls outside the acceptance and thereby reducing $\mathcal{J}_n$.
For example, $\lambda_1 = 0$, $\lambda_2 = \sqrt{2/\pi} \approx 0.80$, and $\lambda_3 = 3/\sqrt{2\pi} \approx 1.20$ (see Appendix~\ref{app:small-sigma} for the general derivation).
All bulk terms cancel in the ratio, yielding
\eq{\label{eq:k4k2-plateau}
\frac{\kappa_4}{\kappa_2} \xrightarrow{\sigma_\eta \to 0} \frac{4\lambda_2 - 6\lambda_3 + 3\lambda_4}{\lambda_2} = -\frac{1}{2} + \frac{9}{\pi}\arcsin\frac{1}{3} \approx 0.474,
}
independent of $\alpha$.
This explains the plateau observed in Fig.~\ref{fig:k2k4-coord} for small $\sigma_\eta$.
Applying the same analysis to $\kappa_6/\kappa_2$ (see Appendix~\ref{app:small-sigma}), one obtains
\eq{\label{eq:k6k2-plateau}
\begin{split}
\frac{\kappa_6}{\kappa_2}
&\xrightarrow{\sigma_\eta \to 0}
\frac{16\lambda_2 - 75\lambda_3 + 150\lambda_4 - 135\lambda_5 + 45\lambda_6}{\lambda_2} \\
&\approx -0.025,
\end{split}
}
a small negative value.
This is a noteworthy consequence of local conservation: even without any critical or medium effects, $\kappa_6/\kappa_2$ is driven to negative values in the small-$\sigma_\eta$ limit.
Since a negative $\kappa_6$ has been proposed as a signature of chiral criticality~\cite{Friman:2011pf}, this finding demonstrates that the conservation baseline must be carefully established before attributing any sign change to critical physics.
The effect is clearly visible in the right panel of Fig.~\ref{fig:k2k4-coord}: the small-$\sigma_\eta$ curves for $\kappa_6/\kappa_2$ form a broad plateau near zero that dips below it, in contrast to the pronounced minima seen for larger $\sigma_\eta$.

\subsection{Comparison with other approaches}

\subsubsection{Thermal-FIST $V_c$ approach}

Figure~\ref{fig:k2k4-coord} also compares the Gaussian local-conservation approach with the $V_c$ approach of Refs.~\cite{Vovchenko:2018fiy,Vovchenko:2019kes,Vovchenko:2024pvk}, in which exact baryon conservation is imposed within a sharp spatial rapidity window $V_c = k\,dV/dy$ (dashed lines). The correspondence between the two prescriptions is established by matching $\kappa_2/\Nacc$ at small $\alpha$, yielding $\sigma_\eta \simeq 0.40, 0.78, 1.20, 2.02$ for $k = 1, 2, 3, 5$, respectively. The two approaches agree at small $\alpha$ but differ qualitatively at larger $\alpha$: the $V_c$ approach drops to zero as soon as the acceptance exceeds the conservation window, since no balancing charge is available outside $V_c$, whereas the Gaussian kernel smoothly interpolates between the local and global regimes. The difference is more pronounced in the higher-order ratios $\kappa_4/\kappa_2$ and $\kappa_6/\kappa_2$, where the sharp cutoff of the $V_c$ approach does not reproduce the characteristic plateau structure predicted by the Gaussian kernel. 

It should be noted that experimental measurements are performed at midrapidity, typically $|y| \lesssim 0.5$, which corresponds to $\alpha \lesssim 0.1$ at LHC energies. Furthermore, since only protons are measured, rather than all baryons, and within a finite $p_T$ acceptance, the effective $\alpha$ is shifted to even smaller values, where the $V_c$ approach remains quantitatively accurate.
For instance, in Ref.~\cite{Braun-Munzinger:2020jbk}, the value of $\alpha$ for net-proton cumulants in Pb--Pb collisions was estimated at $\alpha \approx 0.025$. 
Nevertheless, the results in Fig.~\ref{fig:k2k4-coord} illustrate the advantage of the Gaussian cluster ansatz in providing a physically meaningful description over the full range of acceptances.

The Gaussian local conservation developed here improves upon the $V_c$ approach for the description of net-baryon (net-proton) cumulants.
However, one practical advantage of the $V_c$ approach is that it is implemented as a full hadron resonance gas Monte Carlo event generator in \texttt{Thermal-FIST}~\cite{Vovchenko:2022syc}. 
This makes it particularly valuable for  phenomenological applications beyond the baryon number fluctuations considered here, for instance, when resonance decay feeddown and the associated daughter-particle correlations need to be incorporated explicitly.

\subsubsection{Diffusion master equation}

As a second benchmark, we compare our static Gaussian local-conservation model to the diffusion master equation approach of Ref.~\cite{Sakaida:2014pya}. In that framework, the fireball is modeled as a finite interval in spatial rapidity with reflecting boundaries, and the time evolution of conserved-charge fluctuations is generated by diffusion in rapidity space. The initial state is constrained by global charge conservation, i.e. the net conserved charge in the full system does not fluctuate event by event.

In the diffusion picture, each balancing charge is propagated by a one-body Gaussian kernel of width $d$, where $d$ is the diffusion length accumulated during the hadronic stage. If a balancing pair is created locally and its two members subsequently diffuse independently from the same source point, the corresponding two-body relative kernel is Gaussian with variance $2d^2$. Matching this kernel to our ansatz $\varkappa_2(\eta_1,\eta_2)\propto \exp[-(\eta_1-\eta_2)^2/(2\sigma_\eta^2)]$ gives $\sigma_\eta = \sqrt{2}\,d$. Writing the diffusion length in ballistic regime as $d = \tau\,\eta_{\rm tot}$ and using $\eta_{\rm tot}=2\eta_{\rm max}$, one obtains
\begin{equation}
 \sigma_\eta = 2\sqrt{2}\,\tau\,\eta_{\rm max},
\end{equation}
where $\tau \equiv d/\eta_{\rm tot}$ is the dimensionless diffusion variable (denoted $T$ in Ref.~\cite{Sakaida:2014pya}).

For the comparison to higher-order cumulants, we use the diffusion-master-equation initial condition in which the total particle number is allowed to fluctuate with Poisson statistics. This is the natural choice for comparison to the ideal-gas baseline, where global charge conservation constrains the net charge, but not the total charged-particle number, and the total multiplicity follows Poissonian fluctuations~\cite{Begun:2004gs}. With this choice, we find exact agreement between the diffusion-master-equation results and the present Gaussian model for $\kappa_2/\Nacc$, $\kappa_4/\kappa_2$, and $\kappa_6/\kappa_2$.

\begin{figure*}[t]
    \centering
    \includegraphics[width=0.98\linewidth]{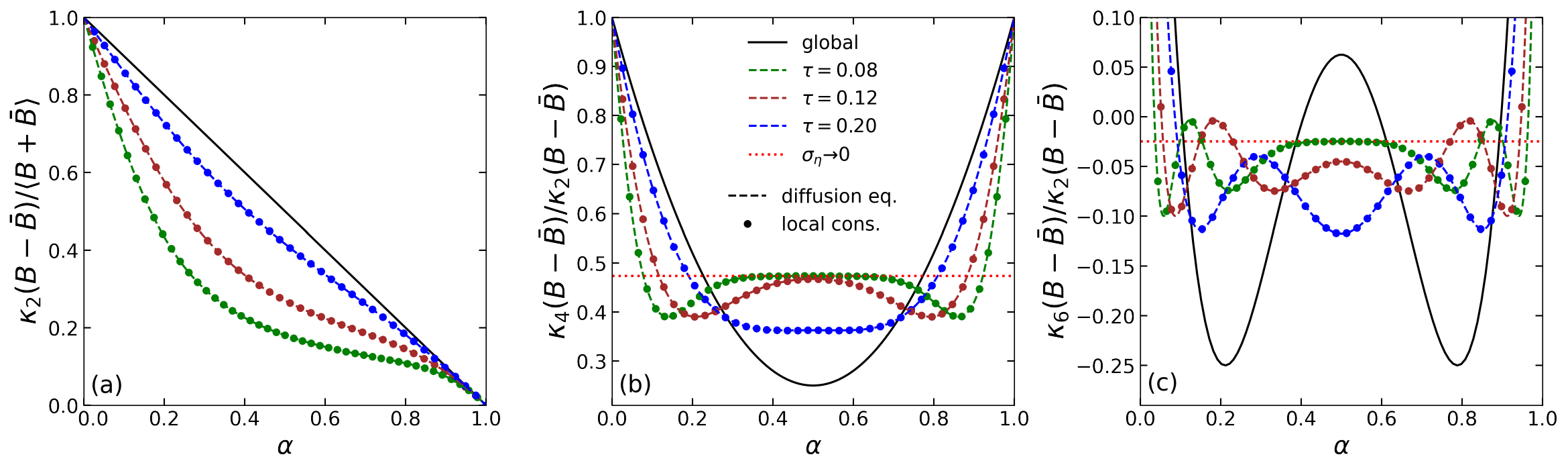}
    \caption{
Comparison of the Gaussian local-conservation approach (black dashed/dash-dotted lines, with $\sigma_\eta = 2\sqrt{2}\,\tau\,\eta_{\rm max}$) with the diffusion master equation of Ref.~\cite{Sakaida:2014pya} (colored dotted lines) for $\kappa_2/\Nacc$~(a), $\kappa_4/\kappa_2$~(b), and $\kappa_6/\kappa_2$~(c) at $\mu_B = 0$.
The solid black line shows the global conservation limit.
The diffusion results are evaluated with the initial condition in which the total particle number fluctuates with Poisson statistics.
With this choice, the Gaussian and diffusion-master-equation results agree exactly for all three cumulant ratios shown.}
    \label{fig:diffusion}
\end{figure*}

An important consequence of this exact agreement is that, in the ideal-gas limit and for the coordinate-space cumulants considered here, the Gaussian local-conservation framework provides an effective static representation of the diffusive broadening generated during the hadronic stage.
In other words, once the width is identified through $\sigma_\eta=\sqrt{2}\,d$, the final-state cumulants obtained from the diffusion master equation can be reproduced entirely in terms of a Gaussian local-conservation kernel.
This gives a simple interpretation of $\sigma_\eta$: in the ideal-gas limit, it encodes the cumulative effect of hadronic diffusion on local baryon-conservation correlations (see also Appendix~\ref{sec-dissipation} for the formal connection between the two frameworks).
We note that the matching relies on the Poissonian total-number initial condition discussed above, which is the one naturally predicted by the canonical ideal gas~\cite{Begun:2004gs}.
Reference~\cite{Sakaida:2014pya} also considers an alternative initial condition with vanishing total-number fluctuations, for which the agreement would not hold.

\subsubsection{Correlated sampling model}

It is also instructive to compare to the correlated sampling model of Ref.~\cite{Braun-Munzinger:2023gsd}. In that approach, full-space multiplicities are sampled from a canonical ensemble distribution, while the rapidities of balancing baryon-antibaryon pairs are generated with a finite rapidity-correlation length. A practical advantage of this construction is that it is naturally implemented as a Monte Carlo procedure and can be combined with essentially arbitrary single-particle rapidity distributions. For second-order cumulants, the behavior of the correlated sampling model agrees well with both the diffusion and Gaussian local-conservation approaches once the parameters are matched appropriately.

For higher-order cumulants, however, the available comparison suggests a qualitatively different behavior. The preliminary results for $\kappa_4/\kappa_2$, reported in Ref.~\cite{Arslandok:2025ntq}~(Fig. 5), indicate that, as the correlations become more local, the ratio tends to move back toward the Poisson limit of unity at a fixed acceptance fraction $\alpha$.
This appears to differ from the behavior of the present Gaussian-cluster model, where more local conservation leads instead to the broad sub-Poisson plateau discussed above. A possible origin of this difference is the different correlation content of the two approaches: the correlated-sampling framework is formulated in a way that can treat unlike-sign or like-sign correlations, but the published implementation~\cite{Braun-Munzinger:2023gsd} does not provide an explicit simultaneous construction of the full set of local $B\bar{B}$, $BB$, and $\bar{B}\bar{B}$ correlations.
This model difference could contribute to the different higher-order behavior, though a definitive conclusion would require a more detailed comparison than what is currently available.
A systematic investigation of higher-order cumulants in the correlated sampling model would therefore be very interesting.

\subsection{Net-proton cumulants in O--O collisions at the LHC}
\label{sec-res-oo}

We now turn to experimentally relevant observables. 
We focus the present analysis on net-proton cumulants in O--O collisions at $\sNN = 5.36$~TeV, where high-statistics measurements from LHC Run~3 are imminent.
The calculations use the blast-wave model~\cite{Vovchenko:2019pjl} to convert coordinate-space correlators into momentum-space distributions, with parameters obtained from Pb--Pb at 5.02 TeV collisions and extrapolated to the multiplicities of O--O collisions at 5.36 TeV.
We use the same $\eta_{\rm max} = 5.1$ as in the coordinate-space analysis and a proton-to-baryon ratio $q = 0.33$ after strong and electromagnetic decays, computed using the canonical statistical model within \texttt{Thermal-FIST}~\cite{Vovchenko:2019kes}.
We note that the blast-wave parameters are extrapolated from Pb--Pb and have not been tuned to O--O data; similarly, $q$ depends on the thermal model composition and could vary at the few-percent level depending on the freeze-out conditions.
These uncertainties are expected to be subdominant compared to the spread between conservation scenarios but should be kept in mind when comparing the predictions with data.
The acceptance probability $p(\eta)$ incorporates both the pseudorapidity cut $|\tilde{\eta}| < \tilde{\eta}_{\rm cut}$ and the transverse-momentum window $0.5 < p_T < 1.5$~GeV/$c$, corresponding to typical ALICE proton reconstruction conditions.
We consider three scenarios for baryon conservation: global conservation, local conservation with $\sigma_\eta = 0.78$ (the value extracted from second-order net-proton cumulants in Pb--Pb collisions in Ref.~\cite{Vovchenko:2024pvk}), and the ultra-local limit $\sigma_\eta \to 0$.\footnote{In contrast to coordinate-space cumulants, which vanish in the ultra-local limit $\sigma_\eta \to 0$, the momentum-space net-proton cumulants remain finite due to thermal smearing and omission of neutrons.}

\begin{figure*}
    \centering
    \includegraphics[width=\linewidth]{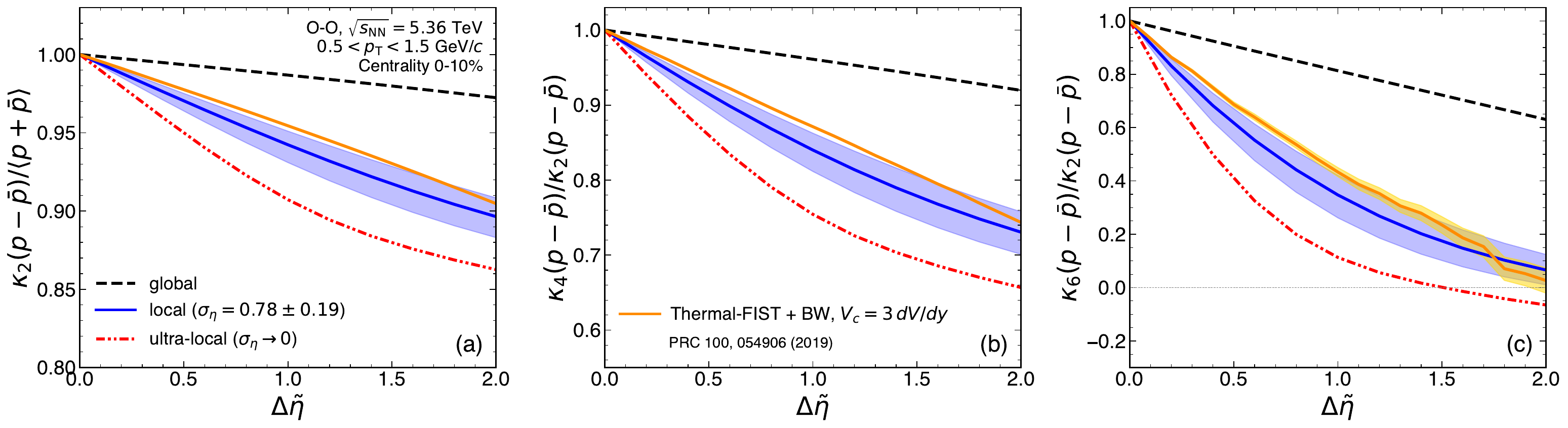}
    \caption{Momentum-space net-proton cumulant ratios $\kappa_2(p{-}\bar{p})/\mean{p{+}\bar{p}}$ (left), $\kappa_4(p{-}\bar{p})/\kappa_2(p{-}\bar{p})$ (middle), and $\kappa_6(p{-}\bar{p})/\kappa_2(p{-}\bar{p})$ (right) as a function of the pseudorapidity acceptance $\Delta\tilde{\eta}$ in $0$--$10\%$ central O--O collisions at $\sNN = 5.36$~TeV.
    Results are shown for global baryon conservation (dashed black), local conservation with $\sigma_\eta = 0.78\pm0.19$ (solid blue with band), and the ultra-local limit $\sigma_\eta \to 0$ (dash-dotted red).
    The orange curve with band shows the \texttt{Thermal-FIST}~$V_c$ approach with $V_c = 3\,dV/dy$~\cite{Vovchenko:2019kes,Vovchenko:2019pjl} for comparison, where the band corresponds to the statistical uncertainty.}
    \label{fig:mom-space-k2k1}
    \label{fig:mom-space-k4k6}
\end{figure*}

Figure~\ref{fig:mom-space-k2k1} shows the momentum-space net-proton cumulant ratios as a function of the pseudorapidity acceptance $\Delta\tilde{\eta}$.
The normalized variance $\kappa_2/\mean{p{+}\bar{p}}$ (left panel) starts from the Skellam baseline of unity at vanishing acceptance and decreases monotonically as $\Delta\tilde{\eta}$ grows, reflecting the buildup of anticorrelations induced by baryon conservation.
The global conservation scenario produces the weakest suppression because the balancing charge is distributed uniformly over the full fireball.
Local conservation with $\sigma_\eta = 0.78$ gives a noticeably steeper suppression, while the ultra-local limit gives the strongest suppression.
At the full ALICE acceptance ($|\tilde{\eta}| < 0.8$, i.e., $\Delta\tilde{\eta} = 1.6$), the three scenarios are sufficiently different to provide a measurable discriminator between global and local conservation in Run~3 data.

The higher-order ratios $\kappa_4/\kappa_2$ and $\kappa_6/\kappa_2$ (middle and right panels of Fig.~\ref{fig:mom-space-k2k1}) are even more sensitive.
Both ratios start at the Skellam value of unity and decrease with increasing acceptance, with the ordering global $>$ local $>$ ultra-local.
At the experimentally relevant ALICE acceptance $\Delta\tilde{\eta} = 1.6$, $\kappa_4/\kappa_2$ ranges from $\approx 0.94$ (global) to $\approx 0.77$~(local $\sigma_\eta = 0.78$).
The sixth-order ratio varies more strongly: at the same acceptance, $\kappa_6/\kappa_2$ varies from $\approx 0.70$ (global) to $\approx 0.15$ (local) to $\approx -0.01$ (ultra-local).
The possibility of $\kappa_6/\kappa_2 < 0$ from local conservation alone is notable, since a negative $\kappa_6$ in full-space has been suggested as a signature of chiral criticality~\cite{Friman:2011pf}.
Our results show that the local conservation baseline must be carefully accounted for before such an interpretation can be made for experimental data in restricted acceptance.

Figure~\ref{fig:mom-space-k2k1} also includes results from the \texttt{Thermal-FIST}~$V_c$ approach (orange) with $V_c = 3\,dV/dy$~\cite{Vovchenko:2019kes}, the value used in the analysis of ALICE Pb--Pb data~\cite{Vovchenko:2024pvk}.
The $V_c$ curve agrees with the Gaussian local conservation result within the $\sigma_\eta = 0.78\pm0.19$ band over most of the acceptance, but the shape of the $\Delta\tilde{\eta}$ dependence is different: the $V_c$ curve is nearly linear in $\Delta\tilde{\eta}$, a consequence of the box-like correlation profile of the $V_c$ approach, while the Gaussian curve is concave.
At small $\Delta\tilde{\eta}$, the $V_c$ value sits slightly above the Gaussian (less suppression) because the two prescriptions are matched to the same Pb--Pb data at larger acceptance, where $\sigma_\eta = 0.78$ and $V_c = 3\,dV/dy$ were extracted.
For $\kappa_6/\kappa_2$ at the largest $\Delta\tilde{\eta}$, the $V_c$ curve drops slightly below the Gaussian band toward the ultra-local limit.
Within the ALICE acceptance ($\Delta\tilde{\eta}\lesssim 1.6$) the two prescriptions are interchangeable. 
However, the Gaussian formulation extends to larger acceptances and avoids the linear box-like behavior at small $\Delta\tilde{\eta}$.

\subsection{Centrality dependence}
\label{sec-res-centrality}

\begin{figure*}
    \centering
    \includegraphics[width=\linewidth]{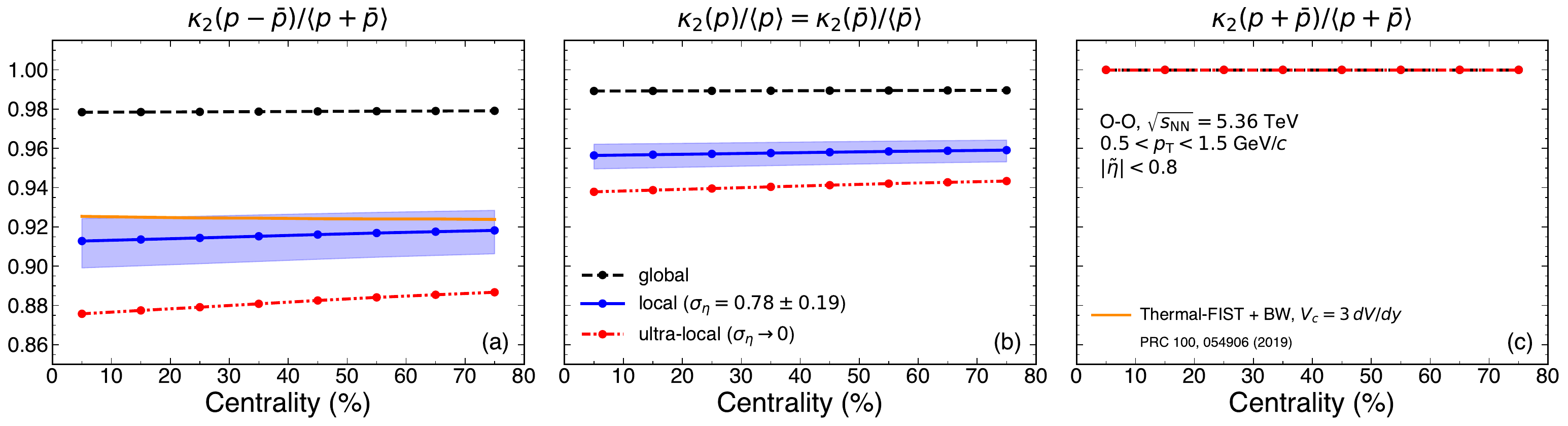}
    \caption{Centrality dependence of second-order proton cumulants in O--O collisions at $\sNN = 5.36$~TeV for $|\tilde{\eta}| < 0.8$ and $0.5 < p_T < 1.5$~GeV/$c$:
    $\kappa_2(p-\bar{p})/\mean{p+\bar{p}}$ (left), $\kappa_2(p)/\mean{p} = \kappa_2(\bar{p})/\mean{\bar{p}}$ (middle), and $\kappa_2(p+\bar{p})/\mean{p+\bar{p}}$ (right).
    Curves and bands as in Fig.~\ref{fig:mom-space-k2k1}, including the \texttt{Thermal-FIST}~$V_c$ approach (orange).
    }
    \label{fig:mom-space-k2-cent}
\end{figure*}

\begin{figure*}
    \centering
    \includegraphics[width=0.49\linewidth]{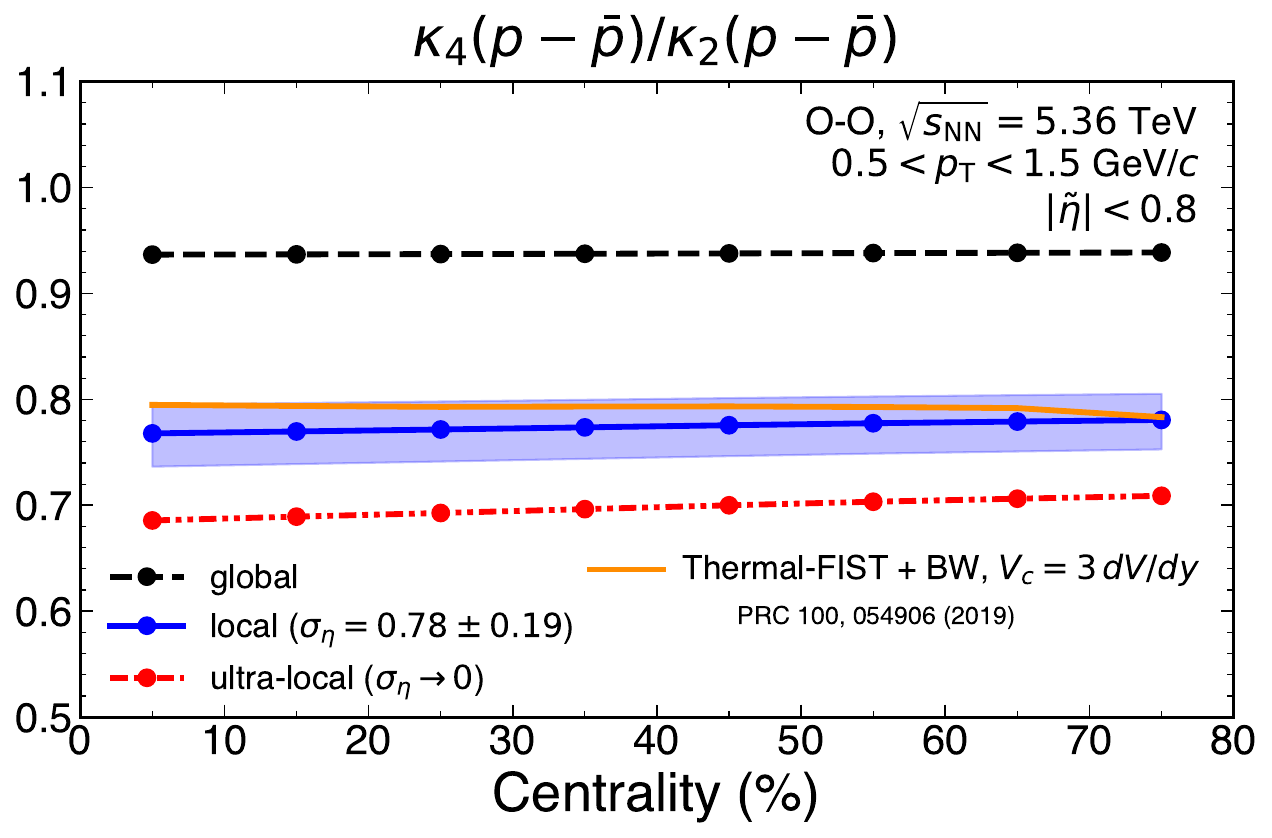}
    \includegraphics[width=0.49\linewidth]{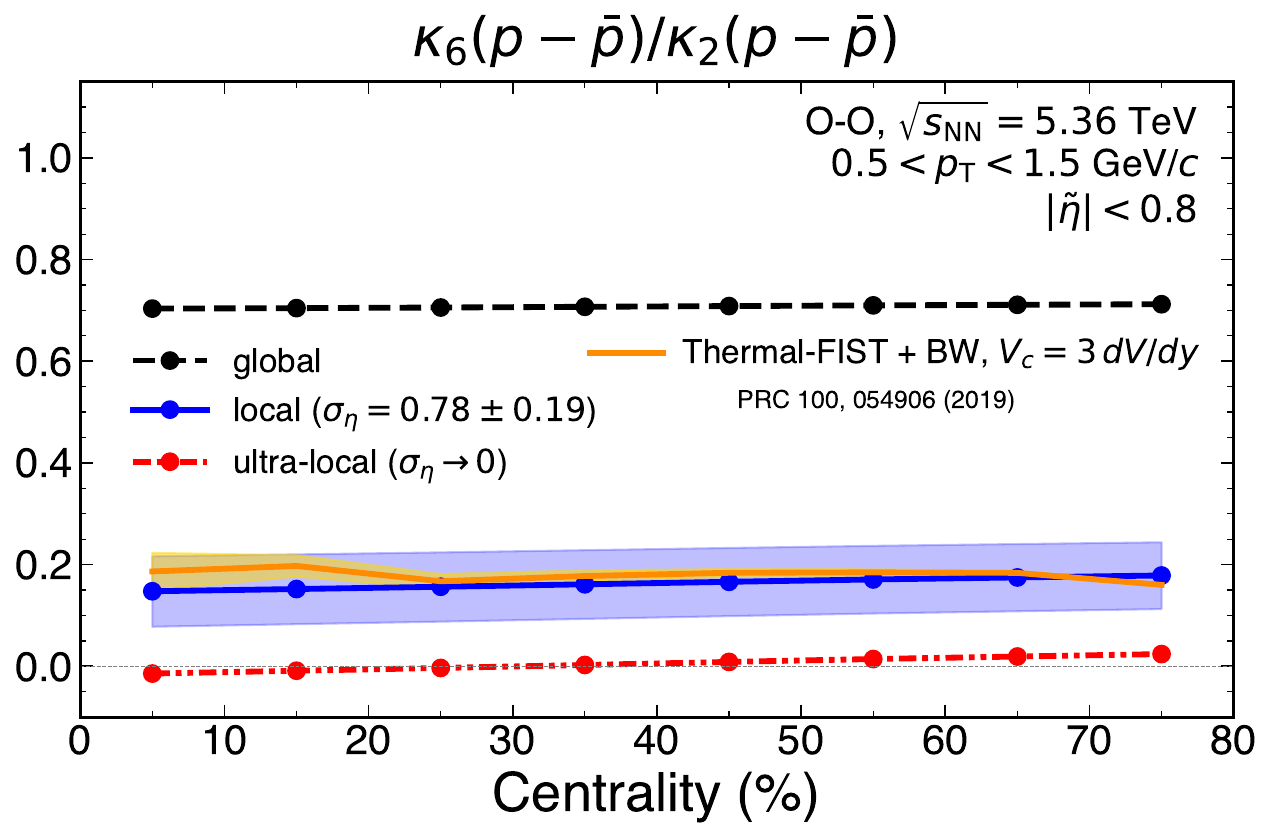}
    \caption{Centrality dependence of the net-proton cumulant ratios $\kappa_4/\kappa_2$ (left) and $\kappa_6/\kappa_2$ (right) in O--O collisions at $\sNN = 5.36$~TeV for $|\tilde{\eta}| < 0.8$ and $0.5 < p_T < 1.5$~GeV/$c$.
    Curves and bands as in Fig.~\ref{fig:mom-space-k2k1}, including the \texttt{Thermal-FIST}~$V_c$ approach (orange).}
    \label{fig:mom-space-k4k6-cent}
\end{figure*}

Figures~\ref{fig:mom-space-k2-cent} and~\ref{fig:mom-space-k4k6-cent} show the centrality dependence of the proton cumulants for the standard acceptance $|\tilde{\eta}| < 0.8$ and $0.5 < p_T < 1.5$~GeV/$c$.
Figure~\ref{fig:mom-space-k2-cent} displays three complementary second-order measures: the net-proton variance $\kappa_2(p-\bar{p})/\mean{p+\bar{p}}$ (left panel), the same-sign proton variance $\kappa_2(p)/\mean{p} = \kappa_2(\bar{p})/\mean{\bar{p}}$ (middle panel), and the summed $\kappa_2(p+\bar{p})/\mean{p+\bar{p}}$ (right panel).
All three measures show only a very weak centrality dependence, a consequence of the fact that both the cumulants and the mean multiplicities scale approximately with the number of participants, so that their ratios are nearly constant.
The net-proton variance exhibits the clearest separation between scenarios ($\approx 0.98$ for global, $\approx 0.92$ for local, $\approx 0.88$ for ultra-local), making it the most discriminating observable among these three.
The same-sign variance shows a similar, somewhat smaller separation.
The summed variance $\kappa_2(p+\bar{p})/\mean{p+\bar{p}}$ is exactly unity for all three scenarios: baryon conservation affects the net-baryon distribution but not the total baryon-plus-antibaryon distribution, since the latter is insensitive to the sign of the baryon number carried by each particle.
This provides a clean experimental cross-check: a deviation of $\kappa_2(p+\bar{p})/\mean{p+\bar{p}}$ from unity would indicate correlations beyond the ideal gas with conservation laws, e.g., from baryon annihilation, interactions, or volume fluctuations.
The \texttt{Thermal-FIST}~$V_c$ result (orange) agrees with the Gaussian local conservation result across centralities for $\kappa_2(p-\bar{p})/\mean{p+\bar{p}}$, with deviations within the $\sigma_\eta = 0.78\pm0.19$ band.

The higher-order ratios in Fig.~\ref{fig:mom-space-k4k6-cent} exhibit similarly weak centrality dependence.
The $\kappa_4/\kappa_2$ ratio takes values of $\approx 0.94$ (global), $\approx 0.77$ (local), and $\approx 0.70$ (ultra-local), with the separation roughly constant across centralities.
The $\kappa_6/\kappa_2$ ratio shows the largest discrimination: $\approx 0.71$ (global), $\approx 0.17$ (local), and $\approx 0.00$ (ultra-local).
In general, the ultra-local scenario provides a lower bound on all the considered cumulant ratios if (local) baryon conservation is the only effect included.

\subsection{Predictions for Pb--Pb collisions}
\label{sec-res-PbPb}

\begin{figure*}
    \centering   \includegraphics[width=\linewidth]{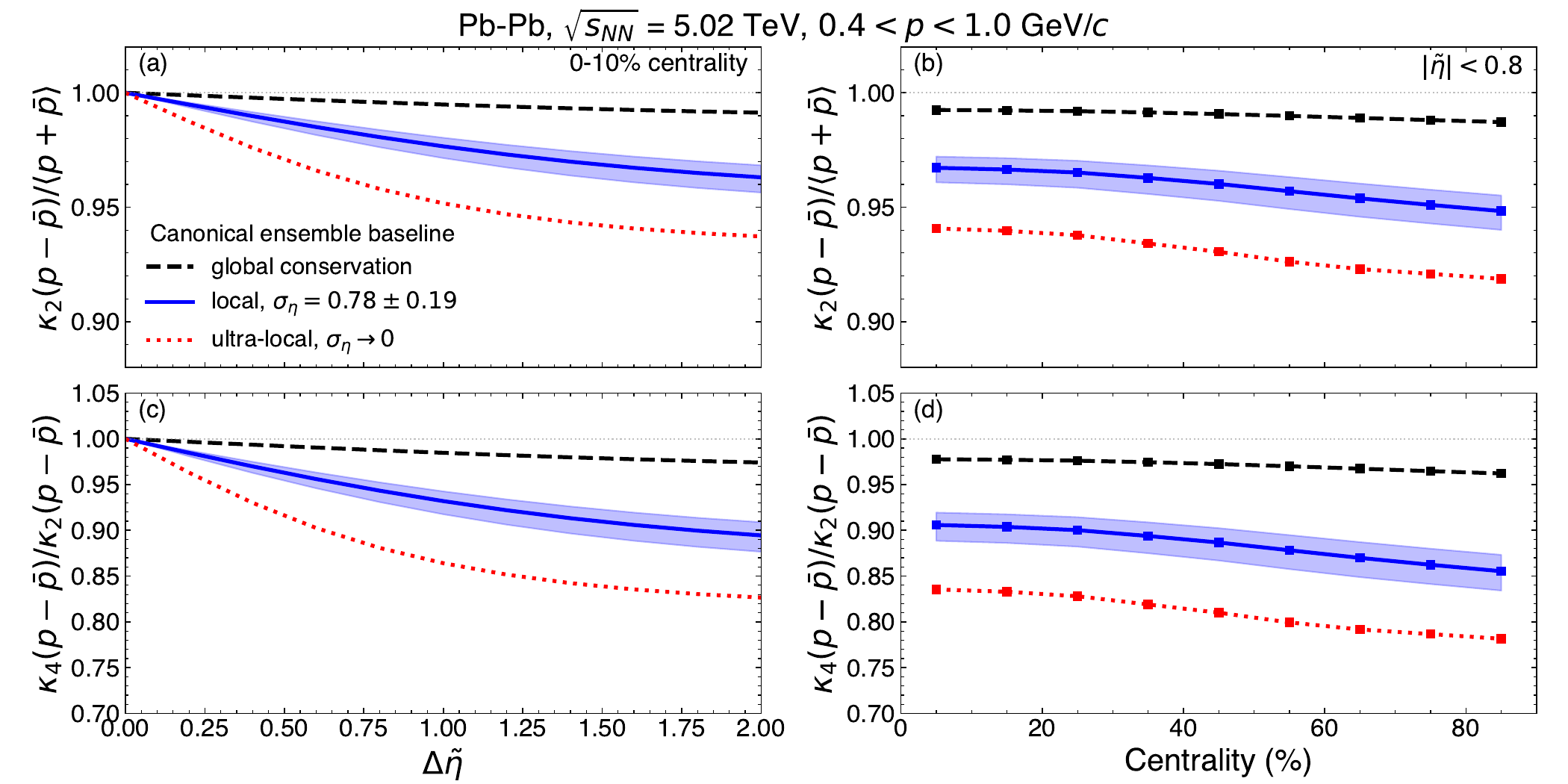}
    \caption{Predictions for net-proton cumulant ratios $\kappa_2(p{-}\bar{p})/\mean{p{+}\bar{p}}$ (top) and $\kappa_4(p{-}\bar{p})/\kappa_2(p{-}\bar{p})$ (bottom) in Pb--Pb collisions at $\sNN = 5.02$~TeV with $0.4 < p < 1.0$~GeV/$c$.
    Left panels show the dependence on the pseudorapidity acceptance $\Delta\eta$ in $0$--$10\%$ centrality; right panels show the centrality dependence at $|\eta| < 0.8$.
    Curves as in Fig.~\ref{fig:mom-space-k2k1}.}
    \label{fig:PbPb-predictions}
\end{figure*}

Figure~\ref{fig:PbPb-predictions} shows predictions for net-proton cumulants in Pb--Pb collisions at $\sNN = 5.02$~TeV, using blast-wave parameters extracted from published ALICE spectra and the proton acceptance $0.4 < p < 1.0$~GeV/$c$, $|\eta| < 0.8$ employed in the preliminary ALICE measurement~\cite{Fokin:2025hbv}.
The conservation effects in Pb--Pb are smaller than in O--O primarily because the narrower momentum cut $0.4 < p < 1.0$~GeV/$c$ selects a smaller fraction of all baryons, reducing the effective acceptance fraction~$\alpha$.
Nevertheless, the three conservation scenarios remain clearly separated, particularly for $\kappa_4/\kappa_2$.
At the standard ALICE acceptance ($|\eta| < 0.8$) in $0$--$10\%$ central collisions, the normalized variance $\kappa_2/\mean{p{+}\bar{p}}$ ranges from $\approx 0.99$ (global) to $\approx 0.97$ (local) to $\approx 0.95$ (ultra-local), while $\kappa_4/\kappa_2$ spans from $\approx 0.98$ (global) to $\approx 0.90$ (local) to $\approx 0.84$ (ultra-local).
The centrality dependence (right panels) shows a gradual increase of the suppression toward peripheral collisions, driven by the decreasing fireball size and correspondingly larger acceptance fraction.

\begin{figure*}
    \centering
    \includegraphics[width=\linewidth]{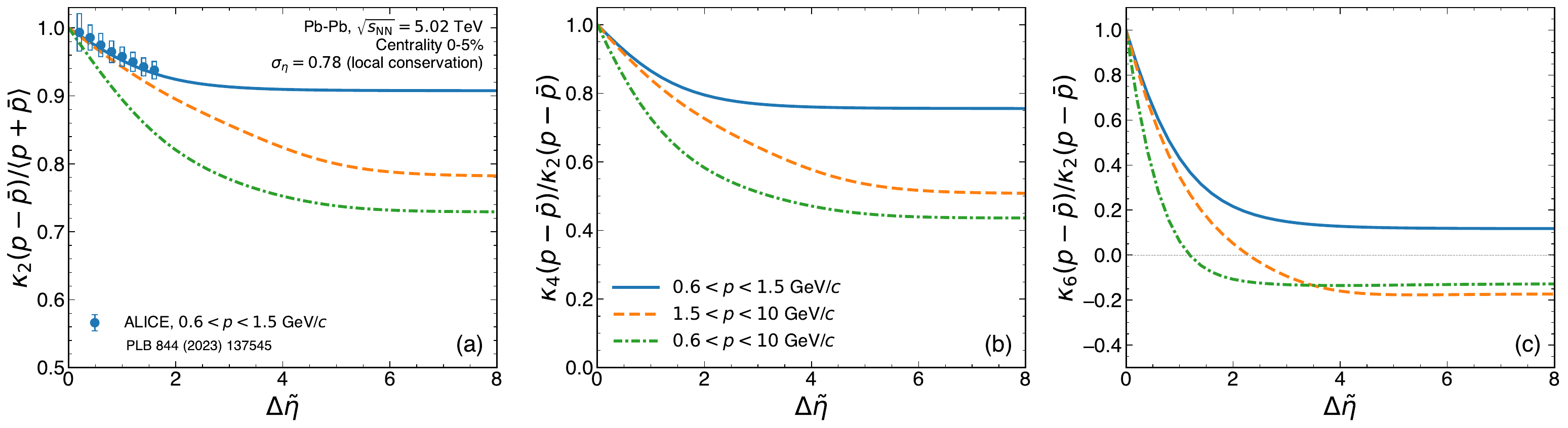}
    \caption{Net-proton cumulant ratios $\kappa_2(p{-}\bar{p})/\mean{p{+}\bar{p}}$ (left), $\kappa_4(p{-}\bar{p})/\kappa_2(p{-}\bar{p})$ (middle), and $\kappa_6(p{-}\bar{p})/\kappa_2(p{-}\bar{p})$ (right) in $0$--$5\%$ central Pb--Pb collisions at $\sNN = 5.02$~TeV, for local conservation with $\sigma_\eta = 0.78$, shown as a function of $\Delta\tilde{\eta}$ up to the projected ALICE 3 coverage~\cite{ALICE:2022wwr,Dainese:2925455}.
    Three momentum windows are compared: the current ALICE acceptance $0.6 < p < 1.5$~GeV/$c$ (solid blue), the high-$p$ extension $1.5 < p < 10$~GeV/$c$ (dashed orange), and the combined $0.6 < p < 10$~GeV/$c$ accessible with ALICE 3 (dash-dotted green).
    The ALICE Pb--Pb data from Ref.~\cite{ALICE:2022xpf} are shown in the left panel.}
    \label{fig:ALICE3-projections}
\end{figure*}

As in the O--O case,
we vary $\sigma_\eta = 0.78 \pm 0.19$, corresponding to the uncertainty on the value extracted from second-order net-proton cumulants in Ref.~\cite{Vovchenko:2024pvk}.
These predictions are intended as a non-critical baseline: they quantify the ideal-gas conservation effect that must be subtracted before any deviation can be attributed to QCD medium effects such as the chiral crossover.
We emphasize that all predictions presented here correspond to a fixed fireball volume.
In experiment, event-by-event fluctuations of the collision geometry (centrality) introduce volume fluctuations that modify the higher-order cumulants~\cite{Skokov:2012ds}; for the normalized variance $\kappa_2[p{-}\bar{p}]/\mean{p{+}\bar{p}}$ these corrections cancel, but for $\kappa_4/\kappa_2$ and higher ratios they must be corrected for before comparison with fixed-volume predictions.
The upcoming ALICE measurements are expected to apply such corrections.
While a direct comparison with the preliminary Pb--Pb data reported by ALICE~\cite{Fokin:2025hbv} is premature at this stage, our baseline curves presented here provide the necessary reference against which future published data can be compared.

Finally, the proposed ALICE 3 detector~\cite{ALICE:2022wwr,Dainese:2925455} is expected to extend both the pseudorapidity coverage and momentum range of the experiment.
As a result, the effective acceptance for net-proton fluctuation measurements will be significantly enlarged: a wider $\Delta\tilde\eta$ samples a bigger portion of the fireball at fixed $\eta_{\rm max}$, and a wider momentum window admits more of the proton spectrum at each rapidity.
Figure~\ref{fig:ALICE3-projections} shows the corresponding predictions for central Pb--Pb collisions with $\sigma_\eta = 0.78$, plotted up to $\Delta\tilde\eta = 8$, for three momentum windows: the current ALICE acceptance $0.6 < p < 1.5$~GeV/$c$ (solid blue), the high-$p$ extension $1.5 < p < 10$~GeV/$c$ (dashed orange), and the combined $0.6 < p < 10$~GeV/$c$ accessible with ALICE 3 (dash-dotted green).
The existing ALICE data~\cite{ALICE:2022xpf} for normalized variance $\kappa_2/(p+\bar{p})$ is shown in the left panel, which agrees with our local conservation calculations within uncertainties. 
The qualitative picture is the same as for O--O but with a much larger lever arm: the variance suppression increases with $\Delta\tilde\eta$, and the higher-order ratios depart from their Skellam values much more strongly than at the current ALICE acceptance.
The behavior of $\kappa_6/\kappa_2$ is distinctive: it turns negative once the acceptance is wide enough as a consequence of local baryon conservation only.
Thus, a proper correction for the conservation effects will be essential to extract the possible signal of chiral criticality.

\section{Discussion and outlook}
\label{sec-summary}

We have incorporated the effect of local baryon conservation into higher-order cumulants by 
extending the density correlator formalism.  
We derived explicit expressions for the $n$-point density correlators $\mathcal{C}_n$ up to $n = 6$ in terms of general local conservation kernels $\varkappa_n$.
These expressions hold for any equation of state and provide a systematic framework for incorporating local conservation effects into higher-order cumulant analyses.
We adopted a symmetric Gaussian cluster ansatz for the $n$-point kernels, controlled by a single rapidity width parameter $\sigma_\eta$ shared across all orders. 
Finite-volume boundary conditions, namely periodic and reflecting, were implemented through image sums, yielding a description of the kernels applicable to a finite spatial rapidity domain.

A technical limitation of the density correlator approach is that the correlators $\mathcal{C}_n$ are derived in the thermodynamic (large-volume) limit, where the susceptibilities $\chi_n^B$ are well-defined intensive quantities and the partition into local and balancing contributions is exact.
For central Pb--Pb collisions at the LHC, the baryon-plus-antibaryon yield per unit rapidity at midrapidity is $\sim O(10^2)$~\cite{Braun-Munzinger:2020jbk}, and this approximation is well justified; even for central O--O collisions, where it is $\sim O(10)$, the thermodynamic limit remains a reasonable approximation.
In peripheral collisions or smaller systems the finite-multiplicity corrections could become relevant, and it would be interesting to quantify their magnitude; these corrections are automatically accounted for in the $V_c$ approach~\cite{Vovchenko:2024pvk} and the correlated sampling method~\cite{Poberezhnyuk:2020ayn}, both of which perform exact canonical sampling at finite particle number.

In the ideal hadron gas at $\mu_B = 0$, all even susceptibilities coincide, and the correlators and the resulting cumulants simplify to polynomials in the acceptance-weighted kernel integrals $\mathcal{J}_n$.
The coordinate-space cumulant ratios $\kappa_4/\kappa_2$ and $\kappa_6/\kappa_2$ are symmetric functions of the spatial rapidity acceptance fraction $\alpha$ around $\alpha = 0.5$, approaching the Skellam value of unity at both $\alpha \to 0$ and $\alpha \to 1$.
In the small-$\sigma_\eta$ limit, both ratios develop a characteristic plateau, which we computed and cross-checked analytically.
We also derived the species-decomposed correlators and acceptance cumulants for baryon-only and baryon-antibaryon observables. 
In particular, the effect of conservation laws is weaker on the (anti)baryon-only scaled variance relative to the net-baryon one and vanishes entirely for the total $B{+}\bar{B}$ variance, which remains exactly Poissonian in any acceptance. The latter result provides a clean experimental probe of physics beyond baryon conservation.
We stress that the factorized acceptance treatment used here relies on the ideal-gas property that all local correlations are purely self-correlations; extending it to an interacting system will require a more refined separation of local many-body correlations from balancing contributions when applying the acceptance filter.

Our coordinate-space results are in exact agreement with the diffusion master equation of Ref.~\cite{Sakaida:2014pya} for all cumulant ratios up to $\kappa_6/\kappa_2$, when the Gaussian kernel width is matched to the diffusion variable $\sigma_\eta = 2\sqrt{2}\,\tau\,\eta_{\rm max}$.
This provides a physical interpretation: the Gaussian local conservation formalism describes the same physics as stochastic diffusion of baryon charge in a non-interacting hadron gas, but in a form that is analytically tractable at high orders and straightforward to combine with kinematic acceptance cuts.
We also compared our results to the commonly used correlation volume $V_c$ approach, showing that the two agree quantitatively at small acceptance fractions $\alpha \lesssim 0.05$-$0.1$ (the agreement is broader for lower-order cumulants), which is the regime relevant for midrapidity measurements at the LHC.
The Gaussian kernel extends the description to the full range of $\alpha$ and avoids the box-like correlation profile inherent to the $V_c$ approach.

Using blast-wave parametrizations of the single-particle spectra, we obtained predictions for net-proton cumulants in O--O collisions at $\sNN = 5.36$~TeV and Pb--Pb collisions at $\sNN = 5.02$~TeV, including the dependence on the pseudorapidity acceptance and centrality.
These ideal-gas predictions establish a non-critical baseline: deviations of experimental data from this baseline could signal contributions from the chiral crossover, such as the suppression of $\chi_4^B/\chi_2^B$ or a negative $\chi_6^B$ expected as a remnant of chiral criticality at $\mu_B \approx 0$~\cite{Friman:2011pf}.
Notably, higher-order ratios, especially $\kappa_6/\kappa_2$, are highly sensitive to local conservation and can be driven to small or negative values in restricted acceptance even without any critical effects, underscoring the importance of establishing this baseline before interpreting experimental data.

The present work leaves ample room for extensions.
First, the formalism can be generalized to incorporate multiple conserved charges (such as baryon number, strangeness, and electric charge) thereby allowing a consistent treatment of proton fluctuations at lower energies, or treatment of (net-)charge or strange hadron yields~\cite{ALICE:2024rnr}. 
Second, the ideal-gas baseline can be replaced by a full-fledged QCD equation-of-state-based calculation, where the susceptibilities $\chi_n^B$ are taken from lattice QCD calculations.
Such an extension would require a more careful treatment of the momentum-space acceptance: the binomial single-particle filter still applies, but the different partition-level terms in the correlator now carry distinct susceptibilities $\chi_n^B$ rather than a single overall factor, so one must track which powers of the acceptance function $p(\eta)$ multiply which susceptibility combinations.
A promising avenue is to use the maximum-entropy approach to the thermal freeze-out~\cite{Pradeep:2022eil,Karthein:2025hvl} to reconstruct the baryon-antibaryon factorial cumulants from lattice susceptibilities, which would provide the necessary input for computing the actual effects of the QCD equation of state on the measured cumulants.
Finally, an extension to lower collision energies accessible at RHIC and FAIR will require going beyond the boost-invariant approximation, accounting for the finite net-baryon density ($\mu_B \neq 0$), and treating the rapidity dependence of the single-particle distributions and susceptibilities explicitly.
We look forward to confronting the predictions presented here with upcoming high-precision data from ALICE Run~3 and beyond.

\section*{Acknowledgments}
We thank Anar Rustamov for fruitful discussions on the correlated sampling approach.
V.A.K. and V.V. have been supported by the U.S. Department of Energy, 
Office of Science, Office of Nuclear Physics, Early Career Research Program under Award Number DE-SC0026065.

\appendix

\section{Fifth- and sixth-order density correlators}
\label{app:C56-global}
\label{app:C56-local}

The fifth- and sixth-order density correlators with local conservation kernels read
\begin{widetext}
\eq{&\mathcal{C}_5(\eta_1,\dots,\eta_5) = \chi^B_5\delta_{1,2,3,4,5} - \frac{\chi_5^B}{4!V}\sum\limits_{\sigma \in S_5} \delta_{\sigma_1,\sigma_2,\sigma_3,\sigma_4}\varkappa_2(\eta_{\sigma_1},\eta_{\sigma_5})-\frac{\chi_3^B \chi_4^B}{3!2! \chi^B_2 V}\sum\limits_{\sigma \in S_5}\delta_{\sigma_1,\sigma_2,\sigma_3}\delta_{\sigma_4,\sigma_5}\varkappa_2(\eta_{\sigma_1},\eta_{\sigma_5}) \nonumber \\&+ \frac{1}{2!3!V^2}\left[\chi_5^B+\frac{\chi_3^B\chi_4^B}{\chi_2^B}\right]\sum\limits_{\sigma \in S_5}\delta_{\sigma_1,\sigma_2,\sigma_3}\varkappa_3(\eta_{\sigma_1},\eta_{\sigma_4},\eta_{\sigma_5})+\frac{2\chi^B_3\chi^B_4}{(2!)^3\chi^B_2 V^2}\sum\limits_{\sigma \in S_5}\delta_{\sigma_1,\sigma_2}\delta_{\sigma_3,\sigma_4}\varkappa_3(\eta_{\sigma_1},\eta_{\sigma_4},\eta_{\sigma_5}) \nonumber \\&- \frac{1}{3!2!V^3}\left[\chi^B_5 + 5\frac{\chi_3^B\chi_4^B}{\chi_2^B}\right]\sum\limits_{\sigma \in S_5} \delta_{\sigma_1,\sigma_2}\varkappa_4(\eta_{\sigma_1},\eta_{\sigma_3},\eta_{\sigma_4},\eta_{\sigma_5}) + \frac{4}{V^4}\left[\chi^B_5 + 5\frac{\chi_3^B\chi_4^B}{\chi_2^B}\right] \varkappa_5(\eta_{1},\dots,\eta_{5}) \label{eq:C5-local}}
\eq{&\mathcal{C}_6(\eta_1,\dots,\eta_6) = \chi^B_6 \delta_{1,2,3,4,5,6} - \frac{\chi_6^B}{5!V}\sum\limits_{\sigma \in S_6} \delta_{\sigma_1,\sigma_2,\sigma_3,\sigma_4,\sigma_5} \varkappa_2(\eta_{\sigma_1},\eta_{\sigma_6}) - \frac{\chi^B_3\chi^B_5}{4!2!\chi_2^B V}\sum\limits_{\sigma \in S_6}\delta_{\sigma_1,\sigma_2,\sigma_3,\sigma_4}\delta_{\sigma_5,\sigma_6}\varkappa_2(\eta_{\sigma_1},\eta_{\sigma_6}) \nonumber \\&- \frac{(\chi^B_4)^2}{2!(3!)^2 V\chi^B_2} \sum\limits_{\sigma \in S_6} \delta_{\sigma_1,\sigma_2,\sigma_3}\delta_{\sigma_4,\sigma_5,\sigma_6}\varkappa_2(\eta_{\sigma_1},\eta_{\sigma_6}) + \frac{1}{4!2!V^2}\left[\chi^B_6 + \frac{\chi^B_3\chi^B_5}{\chi^B_2}\right]\sum\limits_{\sigma \in S_6} \delta_{\sigma_1,\sigma_2,\sigma_3,\sigma_4} \varkappa_3(\eta_{\sigma_1},\eta_{\sigma_5},\eta_{\sigma_6}) \nonumber \\&+ \frac{1}{2!\,3!\,V^2}\left[\frac{(\chi^B_4)^2+\chi_3^B\chi_5^B}{\chi^B_2}\right]\sum\limits_{\sigma \in S_6} \delta_{\sigma_1,\sigma_2,\sigma_3}\delta_{\sigma_4,\sigma_5} \varkappa_3(\eta_{\sigma_1},\eta_{\sigma_5},\eta_{\sigma_6})+\frac{1}{3!(2!)^3V^2}\left[\frac{3\chi^B_2(\chi^B_3)^2\chi^B_4 - (\chi^B_3)^4}{(\chi^B_2)^3}\right] \nonumber \\&\times\sum\limits_{\sigma \in S_6}\delta_{\sigma_1,\sigma_2}\delta_{\sigma_3,\sigma_4}\delta_{\sigma_5,\sigma_6} \varkappa_3(\eta_{\sigma_1},\eta_{\sigma_3},\eta_{\sigma_5}) - \frac{1}{(3!)^2V^3}\left[\frac{2(\chi^B_4)^2+3\chi^B_3\chi^B_5+\chi^B_2\chi^B_6}{\chi^B_2}\right]\sum\limits_{\sigma \in S_6}\delta_{\sigma_1,\sigma_2,\sigma_3}\varkappa_4(\{\eta_{\sigma_i}\}_{i=1,4,5,6}) \nonumber \\&+ \frac{1}{(2!)^4V^3}\left[\frac{(\chi^B_3)^4-3\chi^B_2(\chi^B_3)^2\chi^B_4 - 2 (\chi^B_2)^2(\chi^B_4)^2 - 2(\chi^B_2)^2\chi^B_3\chi^B_5}{(\chi^B_2)^3}\right]\sum\limits_{\sigma \in S_6} \delta_{\sigma_1,\sigma_2}\delta_{\sigma_3,\sigma_4} \varkappa_4(\{\eta_{\sigma_i}\}_{i=1,4,5,6}) \nonumber \\&+ \frac{1}{4!2!V^4}\left[\frac{-3(\chi^B_3)^4+9\chi^B_2\chi^B_4(\chi^B_3)^2 + 9 (\chi^B_2)^2\chi^B_5\chi^B_3+(\chi^B_2)^2(8(\chi^B_4)^2+\chi^B_2\chi^B_6)}{(\chi^B_2)^3}\right]\sum\limits_{\sigma \in S_6} \delta_{\sigma_1,\sigma_2}\varkappa_5(\{\eta_{\sigma_i}\}_{i=1,3,4,5,6}) \nonumber \\&-\frac{5}{V^5}\left[\frac{-3(\chi^B_3)^4+9\chi^B_2\chi^B_4(\chi^B_3)^2+9(\chi^B_2)^2\chi^B_5\chi^B_3 + (\chi^B_2)^2(8(\chi^B_4)^2+\chi^B_2\chi^B_6)}{(\chi^B_2)^3}\right] \varkappa_6(\{\eta_{i}\}_{i=1,2,3,4,5,6}) \label{eq:C6-local}}
where the sum $\sum_{\sigma \in S_k}$ runs over all permutations of the index set $\{1, \dots, k\}$.
The global conservation correlators are recovered by setting $\varkappa_n \to 1$ in the above expressions.
\end{widetext}

\section{Analytical two-point kernel via Jacobi theta functions}
\label{jacobi_AP}

The Gaussian two-point kernel $\varkappa_2(\eta_1,\eta_2)$ can be expressed in closed form for both periodic and reflecting boundary conditions using Jacobi theta functions.

\paragraph{Periodic boundary conditions.}
The infinite-volume Gaussian kernel is periodized on $[-\eta_{\rm max}, \eta_{\rm max}]$ by summing over images. Using the transitional invariance on periodicity interval, one can define $\Delta = \eta_1 - \eta_2$ for simplicity.
Applying the Poisson summation formula yields a compact result in terms of the Jacobi theta function $\theta_3$:
\eq{\label{eq:pbc_kernel}
\varkappa_2^{\text{PBC}}(\Delta) = \,\theta_3\!\left(\frac{\pi \Delta}{2\eta_{\rm max}},\; e^{-\pi^2\sigma_\eta^2/(2\eta_{\rm max}^2)}\right)\!,
}
valid for any ratio $\sigma_\eta/\eta_{\rm max}$.

\paragraph{Reflecting boundary conditions.}
For reflecting (hard-wall) boundaries, the periodicity interval is doubled to $4\eta_{\rm max}$ and a mirror image is added. 
The resulting kernel reads
\begin{widetext}
\eq{\label{eq:apbc_kernel}
\varkappa_2^{\text{RBC}}(\eta_1,\eta_2) = \frac{1}{2}\left[\theta_3\!\left(\frac{\pi (\eta_1{-}\eta_2)}{4\eta_{\rm max}},\; e^{-\pi^2\sigma_\eta^2/(8\eta_{\rm max}^2)}\right) + \theta_3\!\left(\frac{\pi (\eta_1{+}\eta_2+2\eta_{\rm max})}{4\eta_{\rm max}},\; e^{-\pi^2\sigma_\eta^2/(8\eta_{\rm max}^2)}\right)\right]\!,
}
\end{widetext}
where the coordinates are defined on $\eta_{1,2} \in (-\eta_{\rm max},\eta_{\rm max})$. Note, that this result can't be longer expressed with $\Delta$ only as function is no longer periodic on the fireball borders.

We did not obtain a simple theta-function representation for the $n$-point kernels with $n \geq 3$.

\section{\texorpdfstring{Common-source representation and small-$\sigma_\eta$ limit of cumulant ratios}{Common-source representation and small-sigmaeta limit of cumulant ratios}}
\label{app:small-sigma}

\subsection{Common-source representation}

The Gaussian-cluster ansatz of Eq.~\eqref{eq:kappa_n} admits a convenient factorized representation.
Defining
\eq{g_\sigma(x) = \frac{1}{\sqrt{\pi}\,\sigma_\eta}\,e^{-x^2/\sigma_\eta^2},}
one can complete the square in the exponent of Eq.~\eqref{eq:kappa_n} to obtain the infinite-volume identity
\begin{equation}
\frac{\varkappa_n(\eta_1,\ldots,\eta_n)}{V^{n-1}}
= \int_{-\infty}^{\infty} dX\,\prod_{i=1}^n g_\sigma(\eta_i-X).
\label{eq:common-source-inf}
\end{equation}
Thus the $n$ coordinates can be viewed as independent Gaussian offsets from a common source position $X$.

In the finite system considered in the main text it is more convenient to fold the source coordinate back into the fireball interval $X\in[-\eta_{\rm max},\eta_{\rm max}]$ and absorb the images into a one-body kernel $g_{\rm BC}(\eta|X)$.
For periodic boundary conditions one has
\begin{equation}
 g_{\rm PBC}(\eta|X)=\sum_{m\in\mathbb Z} g_\sigma(\eta-X+mV),
 \qquad V\equiv 2\eta_{\rm max},
\label{eq:g-pbc}
\end{equation}
while for reflecting boundary conditions,
\begin{equation}
\begin{split}
 g_{\rm RBC}(\eta|X)
 &= \sum_{m\in\mathbb Z}\Big[g_\sigma(\eta-X+2mV) \\
 &\qquad\qquad\quad + g_\sigma(\eta+X+(2m+1)V)\Big].
\end{split}
\label{eq:g-rbc}
\end{equation}
With these definitions, the acceptance integrals of Eq.~\eqref{eq:Jn-def} reduce exactly to
\begin{equation}
\begin{split}
 \mathcal{J}_n &= \int_{-\eta_{\rm max}}^{\eta_{\rm max}} dX\, q_{\rm BC}(X)^n, \\
 q_{\rm BC}(X) &= \int_{-\eta_{\rm max}}^{\eta_{\rm max}} d\eta\, p(\eta)\,g_{\rm BC}(\eta|X).
\end{split}
\label{eq:Jn-1d}
\end{equation}
Equation~\eqref{eq:Jn-1d} is the main practical advantage of the common-source representation: for the Gaussian-cluster ansatz, the $n$-dimensional integral defining $\mathcal{J}_n$ is replaced by a single one-dimensional quadrature over the source coordinate $X$.

For the step-function acceptance used in Sec.~\ref{sec-res},
\begin{equation}
 p(\eta)=\Theta(\eta_{\rm cut}-|\eta|),
\end{equation}
the inner integral can be performed analytically.
For periodic boundary conditions,
\begin{widetext}
\begin{equation}
 q_{\rm PBC}(X)=\frac{1}{2}\sum_{m\in\mathbb Z}\left[\operatorname{erf}\!\left(\frac{\eta_{\rm cut}-X+mV}{\sigma_\eta}\right)-\operatorname{erf}\!\left(\frac{-\eta_{\rm cut}-X+mV}{\sigma_\eta}\right)\right],
\label{eq:q-pbc}
\end{equation}
while for reflecting boundary conditions,
\begin{equation}
\begin{split}
 q_{\rm RBC}(X)=\frac{1}{2}\sum_{m\in\mathbb Z}\Bigg[&\operatorname{erf}\!\left(\frac{\eta_{\rm cut}-X+2mV}{\sigma_\eta}\right)-\operatorname{erf}\!\left(\frac{-\eta_{\rm cut}-X+2mV}{\sigma_\eta}\right) \\
 &+\operatorname{erf}\!\left(\frac{\eta_{\rm cut}+X+(2m+1)V}{\sigma_\eta}\right)-\operatorname{erf}\!\left(\frac{-\eta_{\rm cut}+X+(2m+1)V}{\sigma_\eta}\right)\Bigg].
\end{split}
\label{eq:q-rbc}
\end{equation}
\end{widetext}
The image sums in Eqs.~\eqref{eq:q-pbc} and \eqref{eq:q-rbc} converge rapidly unless $\sigma_\eta$ becomes comparable to the full system size.

\subsection{\texorpdfstring{Small-$\sigma_\eta$ plateau values}{Small-$\sigma_\eta$ plateau values}}

We now derive the plateau values of $\kappa_4/\kappa_2$ and $\kappa_6/\kappa_2$ in the limit of strongly local conservation,
\begin{equation}
 \sigma_\eta \ll \eta_{\rm cut},
 \qquad
 \sigma_\eta \ll \eta_{\rm max}-\eta_{\rm cut},
\end{equation}
for the step-function acceptance $p(\eta)=\Theta(\eta_{\rm cut}-|\eta|)$.
In this regime, the fireball boundaries are parametrically far from the acceptance edges, so the finite-volume kernels reduce locally to their infinite-volume form, and one may use Eq.~\eqref{eq:common-source-inf}.
Then
\begin{equation}
\begin{split}
 \mathcal{J}_n &= \int_{-\infty}^{\infty} dX\, q(X)^n, \\
 q(X) &= \int_{-\eta_{\rm cut}}^{\eta_{\rm cut}} d\eta\, g_\sigma(\eta-X) \\
 &= \frac{1}{2}\left[\operatorname{erf}\!\left(\frac{\eta_{\rm cut}-X}{\sigma_\eta}\right)
 + \operatorname{erf}\!\left(\frac{\eta_{\rm cut}+X}{\sigma_\eta}\right)\right],
\end{split}
\label{eq:Jn-source}
\end{equation}

For small $\sigma_\eta$, $q(X)\approx 1$ deep inside the acceptance and $q(X)\approx 0$ far outside, with transitions confined to boundary layers of width $O(\sigma_\eta)$ near $X=\pm \eta_{\rm cut}$.
Accordingly,
\begin{equation}
 \mathcal{J}_n = 2\eta_{\rm cut} - \lambda_n\sigma_\eta + O(\sigma_\eta^2).
\label{eq:Jn-small-sigma}
\end{equation}
The coefficient $\lambda_n$ measures the expected maximum displacement of $n$ independent Gaussian offsets from the source and is given by
\begin{equation}
 \lambda_n = \sqrt{2}\,\mathbb E[M_n]
 = \sqrt{2}\,n\int_{-\infty}^{\infty}dz\, z\,\phi(z)\,\Phi(z)^{n-1},
\label{eq:lambda-n}
\end{equation}
where $M_n=\max(Z_1,\ldots,Z_n)$ is the maximum of $n$ independent standard normal random variables $Z_i\sim N(0,1)$, $\phi(z) = (2\pi)^{-1/2}\,e^{-z^2/2}$ is the standard normal density, and $\Phi(z) = \int_{-\infty}^z \phi(z')\,dz'$ is its cumulative distribution function.
The integral follows from the order-statistic identity $\mathbb{P}(M_n\leq z) = \Phi(z)^n$, with the factor $n$ accounting for the $n$ equivalent choices of which variable attains the maximum.
The first few values are
\begin{align}
 \lambda_2 &= \sqrt{\frac{2}{\pi}},
 \\
 \lambda_3 &= \frac{3}{\sqrt{2\pi}}, \\
 \lambda_4 &= \frac{3}{\sqrt{2\pi}}\left[1+\frac{2}{\pi}\arcsin\!\left(\frac{1}{3}\right)\right],
 \\
 \lambda_5 &= \frac{5}{2\sqrt{2\pi}}\left[1+\frac{6}{\pi}\arcsin\!\left(\frac{1}{3}\right)\right], \\
 \lambda_6 &=
 \frac{15}{8\sqrt{2\pi}}\left[
 1+\frac{12}{\pi}\arcsin\!\left(\frac{1}{3}\right)+\frac{I_4}{\sqrt{\pi}}
 \right],
 \\
 I_4 &\equiv \int_{-\infty}^{\infty}dz\, e^{-z^2}\,
 \operatorname{erf}^4\!\left(\frac{z}{\sqrt{2}}\right) \nonumber \\
 &\simeq 0.1732.
\label{eq:lambda-6}
\end{align}

Substituting Eq.~\eqref{eq:Jn-small-sigma} into the cumulant ratios, all bulk terms proportional to $2\eta_{\rm cut}$ cancel, since $1-4+6-3=0$ for $\kappa_4$ and $1-16+75-150+135-45=0$ for $\kappa_6$.
One therefore obtains acceptance-independent plateaus.
For $\kappa_4/\kappa_2$,
\begin{equation}
 \begin{split}
 \frac{\kappa_4}{\kappa_2} &\xrightarrow{\sigma_\eta\to 0}
 \frac{4\lambda_2-6\lambda_3+3\lambda_4}{\lambda_2} \\
 &= -\frac{1}{2}+\frac{9}{\pi}\arcsin\!\left(\frac{1}{3}\right)
 \simeq 0.47356.
 \end{split}
\label{eq:k4k2-plateau-app}
\end{equation}
For $\kappa_6/\kappa_2$,
\begin{align}
 \frac{\kappa_6}{\kappa_2} & \xrightarrow{\sigma_\eta\to 0} \frac{16\lambda_2-75\lambda_3+150\lambda_4-135\lambda_5+45\lambda_6}{\lambda_2} \nonumber \\
 & = \frac{31}{16}-\frac{225}{4\pi}\arcsin\!\left(\frac{1}{3}\right)+\frac{675}{16\sqrt{\pi}}\,I_4 
 \simeq -0.02465.
\label{eq:k6k2-plateau-app}
\end{align}
Both plateau values are independent of $\eta_{\rm cut}$ and $\eta_{\rm max}$.
Finally, for the second-order ratio one finds
\begin{equation}
 \frac{\kappa_2}{\Nacc} \xrightarrow{\sigma_\eta\to 0} \frac{\lambda_2\sigma_\eta}{2\eta_{\rm cut}}
 = \frac{\sigma_\eta}{2\eta_{\rm cut}}\sqrt{\frac{2}{\pi}},
\end{equation}
showing that the variance vanishes linearly with $\sigma_\eta$ in the strictly local limit.

\section{Connection with dissipation time}
\label{sec-dissipation}

The model used here is purely static. However, one can interpret a Gaussian kernel with a given correlation width as an averaged ``snapshot'' at a dynamical time $\tau$ with Poisson initial conditions.

According to Ref.~\cite{Sakaida:2017rtj}, one can write a second-order correlation kernel in a form similar to ours, with symmetric and balancing parts:
\eq{C(\Delta, \tau) = \chi(\tau)\delta(\Delta) - \frac{1}{\sqrt{\pi}} \int\limits_{\tau_0}^\tau dt \frac{\dot \chi(t)}{d(t,\tau)} {\rm exp}\left(-\frac{\Delta^2}{4d^2(t,\tau)}\right)}
where $\Delta = \eta_1 - \eta_2$, $\chi(t)$ is a time-dependent susceptibility, and $d(t,\tau)$ is a dissipation length. In general, these two functions are model-dependent. For the ideal-gas case, however, one can use the simple estimate of ballistic linear diffusion, $D(t) = Dt$ with $D = 4\eta_{\rm max}^2$, and take $\chi(t)$ to follow the system temperature $T(t)/T_0$~\cite{Sakaida:2014pya,Sakaida:2017rtj}:
\eq{\begin{split}
    d^2(t,\tau) &= 2\int\limits_{t}^\tau D(t') dt'  = 4\eta^2_{\rm max}(\tau^2 - t^2),
    \\\chi(t) &= \chi_0T(\tau,\tau_0)/T_0 =\chi_0 \frac{\tau_0}{t}
\end{split}}

Note that the form of $d(t,\tau)$ follows from the assumption of linear diffusion in the ideal gas, while $\chi(t)$ is chosen in a simple limiting form. In fact, any function scaling with the hydrodynamic temperature as $\chi(\tau) = \chi_0 T(\tau,\tau_0)/T_0 = \chi_0 (\tau_0/\tau)^{c^2_s}$, with $0<c_s^2\leq1$, gives the same result for $\sigma^2$.

To avoid a normalization mismatch, we use the following definition of $\sigma$:
\eq{\sigma^2(\Delta,\tau) = - \frac{\mean{C(\Delta,\tau)}_\tau}{\mean{C''(\Delta,\tau)}_\tau}\label{eq:sigma2}}
where the time average is defined as
\eq{\mean{X(t)}_\tau = \lim\limits_{\tau_0 \to 0} \frac{1}{(\tau - \tau_0)} \int\limits_{\tau_0}^\tau  dt~ X(t),}
where the limit $\tau_0 \to 0$ is taken for consistency with our model. Effectively, this corresponds to averaging over times long enough for the system to lose memory of the initial condition. In this case, the existence of a constant $\sigma^2$ follows from the theorem that fluctuations in Markovian processes approach a Gaussian form under thermalization (see \cite{gardiner2009stochastic}).

The symmetric part vanishes after time averaging, which means that the background contribution is irrelevant on large time scales $\tau \gg \tau_0$ or $\tau_0/\tau =\varepsilon \to 0$, and therefore $\sigma^2(\tau)$ is unaffected. 
\eq{\begin{split}
    \mean{C_{\rm sym}(\Delta,\tau)}_\tau &= \delta(\Delta)\lim\limits_{\varepsilon\to 0}\frac{\chi_0\varepsilon}{(1-\varepsilon)} \int \limits_{\varepsilon \tau}^{\tau}\frac{dt}{t} \\&=\delta(\Delta)\lim\limits_{\varepsilon \to 0}\frac{\chi_0\varepsilon}{(1-\varepsilon)}{\rm ln}\left[\varepsilon\right] =0
\end{split}}

For the balancing part, one takes the limit $\varepsilon \to 0$, making the result independent of the precise value of $\tau$. For $\Delta \ne 0$, the $O(\varepsilon^n)$ corrections vanish for any finite $n$, since the integrand is exponentially small. The nonzero contribution is therefore again proportional to $\delta(\Delta)$:
\eq{\begin{split}
&\mean{C_{\rm bal}(0,t)}_\tau = \lim\limits_{\varepsilon \to 0} \frac{\chi_0\varepsilon}{\sqrt{\pi}(1-\varepsilon)} \int \limits_{\varepsilon \tau}^\tau d \tau' \int \limits_{\varepsilon \tau}^{\tau'} \frac{t^{-2}dt}{\sqrt{D(\tau^2 - t^2)}}
\\&= \lim\limits_{\varepsilon \to 0} \frac{\chi_0\varepsilon}{\sqrt{\pi}(1-\varepsilon)} \int \limits_{\varepsilon \tau}^\tau dt \frac{\tau -t}{t^2\sqrt{D(\tau^2-t^2)}} = \frac{\chi_0}{\tau \sqrt{\pi D}},
\\& \mean{C''_{\rm bal}(0,t)}_\tau = -\lim\limits_{\varepsilon \to 0} \frac{\chi_0\varepsilon}{2\sqrt{\pi}(1-\varepsilon)} \int \limits_{\varepsilon\tau}^\tau dt \frac{\tau -t}{t^2(D(\tau^2-t^2))^{3/2}} \\&= -\frac{\chi_0}{2\tau^3 \sqrt{\pi} D^{3/2}}
\end{split}}
Using Eq.~\eqref{eq:sigma2}, one obtains
\eq{\sigma^2(\tau) = 8\eta^2_{\rm max}\tau^2}

This result is exact for the second-order cumulant at any finite $\tau > 0$ and has been verified numerically for all cumulant orders considered in this work (see Fig.~\ref{fig:diffusion}). It is a manifestation of the Gaussian nature of the fluctuations in a uniform, non-interacting system once the memory of the initial condition has been erased.

\bibliography{main}

@article{Karthein:2025hvl,
    author = "Karthein, Jamie M. and Rajagopal, Krishna and Pradeep, Maneesha Sushama and Stephanov, Mikhail and Yin, Yi",
    title = "{Quantifying fluctuation signatures of the QCD critical point using maximum entropy freeze-out}",
    eprint = "2508.19237",
    archivePrefix = "arXiv",
    primaryClass = "nucl-th",
    reportNumber = "MIT-CTP/5906",
    doi = "10.1103/9sdb-m9xy",
    journal = "Phys. Rev. D",
    volume = "113",
    number = "7",
    pages = "074010",
    year = "2026"
}

@article{Pratt:2019fbj,
    author = "Pratt, Scott",
    title = "{Calculating $n$-Point Charge Correlations in Evolving Systems}",
    eprint = "1908.01053",
    archivePrefix = "arXiv",
    primaryClass = "nucl-th",
    doi = "10.1103/PhysRevC.101.014914",
    journal = "Phys. Rev. C",
    volume = "101",
    number = "1",
    pages = "014914",
    year = "2020"
}

@article{Koch:2025cog,
    author = "Koch, Volker and Vovchenko, Volodymyr",
    title = "{Exploring the QCD phase diagram through correlations and fluctuations}",
    eprint = "2512.04288",
    archivePrefix = "arXiv",
    primaryClass = "nucl-th",
    month = "12",
    year = "2025"
}

@inproceedings{Arslandok:2025ntq,
    author = "Arslandok, Mesut",
    title = "{Probing bulk dynamics of the QGP with correlations and fluctuations}",
    booktitle = "{31st International Conference on Ultra-relativistic Nucleus-Nucleus Collisions}",
    eprint = "2510.09115",
    archivePrefix = "arXiv",
    primaryClass = "hep-ex",
    month = "10",
    year = "2025"
}

@article{Sakaida:2017rtj,
    author = "Sakaida, Miki and Asakawa, Masayuki and Fujii, Hirotsugu and Kitazawa, Masakiyo",
    title = "{Dynamical evolution of critical fluctuations and its observation in heavy ion collisions}",
    eprint = "1703.08008",
    archivePrefix = "arXiv",
    primaryClass = "nucl-th",
    reportNumber = "J-PARC-TH-0088",
    doi = "10.1103/PhysRevC.95.064905",
    journal = "Phys. Rev. C",
    volume = "95",
    number = "6",
    pages = "064905",
    year = "2017"
}

@article{Sakaida:2014pya,
    author = "Sakaida, Miki and Asakawa, Masayuki and Kitazawa, Masakiyo",
    title = "{Effects of global charge conservation on time evolution of cumulants of conserved charges in relativistic heavy ion collisions}",
    eprint = "1409.6866",
    archivePrefix = "arXiv",
    primaryClass = "nucl-th",
    doi = "10.1103/PhysRevC.90.064911",
    journal = "Phys. Rev. C",
    volume = "90",
    number = "6",
    pages = "064911",
    year = "2014"
}

@book{gardiner2009stochastic,
  title={Stochastic methods},
  author={Gardiner, Crispin},
  volume={4},
  year={2009},
  publisher={Springer Berlin Heidelberg}
}

@inbook{Koch:2008ia,
    author = "Koch, Volker",
    editor = "Stock, R.",
    title = "{Hadronic Fluctuations and Correlations}",
    booktitle = "{Relativistic Heavy Ion Physics}",
    eprint = "0810.2520",
    archivePrefix = "arXiv",
    primaryClass = "nucl-th",
    doi = "10.1007/978-3-642-01539-7_20",
    pages = "626--652",
    year = "2010"
}

@article{Vovchenko:2022syc,
    author = "Vovchenko, Volodymyr",
    title = "{Cooper-Frye sampling with short-range repulsion}",
    eprint = "2208.13693",
    archivePrefix = "arXiv",
    primaryClass = "hep-ph",
    reportNumber = "INT-PUB-22-035",
    doi = "10.1103/PhysRevC.106.064906",
    journal = "Phys. Rev. C",
    volume = "106",
    number = "6",
    pages = "064906",
    year = "2022"
}

@article{Stephanov:1999zu,
	archiveprefix = {arXiv},
	author = {Stephanov, Misha A. and Rajagopal, K. and Shuryak, Edward V.},
	doi = {10.1103/PhysRevD.60.114028},
	eprint = {hep-ph/9903292},
	journal = {Phys. Rev. D},
	pages = {114028},
	reportnumber = {ITP-SB-99-4, MIT-CTP-2834, SUNY-NTG-99-3},
	title = {{Event-by-event fluctuations in heavy ion collisions and the QCD critical point}},
	volume = {60},
	year = {1999}}

@article{Bzdak:2012an,
	archiveprefix = {arXiv},
	author = {Bzdak, Adam and Koch, Volker and Skokov, Vladimir},
	doi = {10.1103/PhysRevC.87.014901},
	eprint = {1203.4529},
	journal = {Phys. Rev. C},
	number = {1},
	pages = {014901},
	primaryclass = {hep-ph},
	reportnumber = {BNL-97063-2012-JA, RBRC-946},
	title = {{Baryon number conservation and the cumulants of the net proton distribution}},
	volume = {87},
	year = {2013}}

@article{Bzdak:2019pkr,
	archiveprefix = {arXiv},
	author = {Bzdak, Adam and Esumi, Shinichi and Koch, Volker and Liao, Jinfeng and Stephanov, Mikhail and Xu, Nu},
	doi = {10.1016/j.physrep.2020.01.005},
	eprint = {1906.00936},
	journal = {Phys. Rept.},
	pages = {1--87},
	primaryclass = {nucl-th},
	title = {{Mapping the Phases of Quantum Chromodynamics with Beam Energy Scan}},
	volume = {853},
	year = {2020}}

@article{Poberezhnyuk:2020ayn,
	archiveprefix = {arXiv},
	author = {Poberezhnyuk, Roman V. and Savchuk, Oleh and Gorenstein, Mark I. and Vovchenko, Volodymyr and Taradiy, Kirill and Begun, Viktor V. and Satarov, Leonid and Steinheimer, Jan and Stoecker, Horst},
	doi = {10.1103/PhysRevC.102.024908},
	eprint = {2004.14358},
	journal = {Phys. Rev. C},
	number = {2},
	pages = {024908},
	primaryclass = {hep-ph},
	title = {{Critical point fluctuations: Finite size and global charge conservation effects}},
	volume = {102},
	year = {2020}}

@article{Vovchenko:2020tsr,
	archiveprefix = {arXiv},
	author = {Vovchenko, Volodymyr and Savchuk, Oleh and Poberezhnyuk, Roman V. and Gorenstein, Mark I. and Koch, Volker},
	doi = {10.1016/j.physletb.2020.135868},
	eprint = {2003.13905},
	journal = {Phys. Lett. B},
	pages = {135868},
	primaryclass = {hep-ph},
	title = {{Connecting fluctuation measurements in heavy-ion collisions with the grand-canonical susceptibilities}},
	volume = {811},
	year = {2020}}

@article{Vovchenko:2021kxx,
	archiveprefix = {arXiv},
	author = {Vovchenko, Volodymyr and Koch, Volker and Shen, Chun},
	doi = {10.1103/PhysRevC.105.014904},
	eprint = {2107.00163},
	journal = {Phys. Rev. C},
	number = {1},
	pages = {014904},
	primaryclass = {hep-ph},
	title = {{Proton number cumulants and correlation functions in Au-Au collisions at sNN=7.7\textendash{}200 GeV from hydrodynamics}},
	volume = {105},
	year = {2022}}

@article{Parra:2025fse,
    author = "Parra, Jonathan and Poberezhniuk, Roman and Koch, Volker and Ratti, Claudia and Vovchenko, Volodymyr",
    title = "{Indications for Freeze-Out of Charge Fluctuations in the Quark-Gluon Plasma at the LHC}",
    eprint = "2504.02085",
    archivePrefix = "arXiv",
    primaryClass = "hep-ph",
    doi = "10.1103/sw74-7hnb",
    journal = "Phys. Rev. Lett.",
    volume = "135",
    number = "24",
    pages = "242302",
    year = "2025"
}

@article{Vovchenko:2024pvk,
    author = "Vovchenko, Volodymyr",
    title = "{Density correlations under global and local charge conservation}",
    eprint = "2409.01397",
    archivePrefix = "arXiv",
    primaryClass = "hep-ph",
    doi = "10.1103/PhysRevC.110.L061902",
    journal = "Phys. Rev. C",
    volume = "110",
    number = "6",
    pages = "L061902",
    year = "2024"
}

@article{HotQCD:2018pds,
    author = "Bazavov, A. and others",
    collaboration = "HotQCD",
    title = "{Chiral crossover in QCD at zero and non-zero chemical potentials}",
    eprint = "1812.08235",
    archivePrefix = "arXiv",
    primaryClass = "hep-lat",
    doi = "10.1016/j.physletb.2019.05.013",
    journal = "Phys. Lett. B",
    volume = "795",
    pages = "15--21",
    year = "2019"
}

@article{Borsanyi:2020fev,
    author = "Borsanyi, Szabolcs and Fodor, Zoltan and Guenther, Jana N. and Kara, Ruben and Katz, Sandor D. and Parotto, Paolo and Pasztor, Attila and Ratti, Claudia and Szabo, Kalman K.",
    title = "{QCD Crossover at Finite Chemical Potential from Lattice Simulations}",
    eprint = "2002.02821",
    archivePrefix = "arXiv",
    primaryClass = "hep-lat",
    doi = "10.1103/PhysRevLett.125.052001",
    journal = "Phys. Rev. Lett.",
    volume = "125",
    number = "5",
    pages = "052001",
    year = "2020"
}

@article{Pisarski:1983ms,
    author = "Pisarski, Robert D. and Wilczek, Frank",
    title = "{Remarks on the Chiral Phase Transition in Chromodynamics}",
    reportNumber = "NSF-ITP-83-152",
    doi = "10.1103/PhysRevD.29.338",
    journal = "Phys. Rev. D",
    volume = "29",
    pages = "338--341",
    year = "1984"
}

@article{HotQCD:2019xnw,
    author = "Ding, H. T. and others",
    collaboration = "HotQCD",
    title = "{Chiral Phase Transition Temperature in ( 2+1 )-Flavor QCD}",
    eprint = "1903.04801",
    archivePrefix = "arXiv",
    primaryClass = "hep-lat",
    doi = "10.1103/PhysRevLett.123.062002",
    journal = "Phys. Rev. Lett.",
    volume = "123",
    number = "6",
    pages = "062002",
    year = "2019"
}

@article{Friman:2011pf,
    author = "Friman, B. and Karsch, F. and Redlich, K. and Skokov, V.",
    title = "{Fluctuations as probe of the QCD phase transition and freeze-out in heavy ion collisions at LHC and RHIC}",
    eprint = "1103.3511",
    archivePrefix = "arXiv",
    primaryClass = "hep-ph",
    reportNumber = "BI-TP-2011-07",
    doi = "10.1140/epjc/s10052-011-1694-2",
    journal = "Eur. Phys. J. C",
    volume = "71",
    pages = "1694",
    year = "2011"
}

@article{ALICE:2022wpn,
    author = "Acharya, Shreyasi and others",
    collaboration = "ALICE",
    title = "{The ALICE experiment: a journey through QCD}",
    eprint = "2211.04384",
    archivePrefix = "arXiv",
    primaryClass = "nucl-ex",
    reportNumber = "CERN-EP-2022-227",
    doi = "10.1140/epjc/s10052-024-12935-y",
    journal = "Eur. Phys. J. C",
    volume = "84",
    number = "8",
    pages = "813",
    year = "2024"
}

@article{Andronic:2017pug,
    author = "Andronic, Anton and Braun-Munzinger, Peter and Redlich, Krzysztof and Stachel, Johanna",
    title = "{Decoding the phase structure of QCD via particle production at high energy}",
    eprint = "1710.09425",
    archivePrefix = "arXiv",
    primaryClass = "nucl-th",
    doi = "10.1038/s41586-018-0491-6",
    journal = "Nature",
    volume = "561",
    number = "7723",
    pages = "321--330",
    year = "2018"
}

@article{Vovchenko:2018fmh,
    author = "Vovchenko, Volodymyr and Gorenstein, Mark I. and Stoecker, Horst",
    title = "{Finite resonance widths influence the thermal-model description of hadron yields}",
    eprint = "1807.02079",
    archivePrefix = "arXiv",
    primaryClass = "nucl-th",
    doi = "10.1103/PhysRevC.98.034906",
    journal = "Phys. Rev. C",
    volume = "98",
    number = "3",
    pages = "034906",
    year = "2018"
}

@article{Begun:2004gs,
    author = "Begun, V. V. and Gazdzicki, M. and Gorenstein, Mark I. and Zozulya, O. S.",
    title = "{Particle number fluctuations in canonical ensemble}",
    eprint = "nucl-th/0404056",
    archivePrefix = "arXiv",
    doi = "10.1103/PhysRevC.70.034901",
    journal = "Phys. Rev. C",
    volume = "70",
    pages = "034901",
    year = "2004"
}

@article{Braun-Munzinger:2020jbk,
    author = "Braun-Munzinger, P. and Friman, B. and Redlich, K. and Rustamov, A. and Stachel, J.",
    title = "{Relativistic nuclear collisions: Establishing a non-critical baseline for fluctuation measurements}",
    eprint = "2007.02463",
    archivePrefix = "arXiv",
    primaryClass = "nucl-th",
    reportNumber = "CERN-TH-2020-116",
    doi = "10.1016/j.nuclphysa.2021.122141",
    journal = "Nucl. Phys. A",
    volume = "1008",
    pages = "122141",
    year = "2021"
}

@article{Vovchenko:2020kwg,
    author = "Vovchenko, Volodymyr and Koch, Volker",
    title = "{Particlization of an interacting hadron resonance gas with global conservation laws for event-by-event fluctuations in heavy-ion collisions}",
    eprint = "2012.09954",
    archivePrefix = "arXiv",
    primaryClass = "hep-ph",
    doi = "10.1103/PhysRevC.103.044903",
    journal = "Phys. Rev. C",
    volume = "103",
    number = "4",
    pages = "044903",
    year = "2021"
}

@article{Castorina:2013mba,
    author = "Castorina, P. and Satz, H.",
    title = "{Causality Constraints on Hadron Production In High Energy Collisions}",
    eprint = "1310.6932",
    archivePrefix = "arXiv",
    primaryClass = "hep-ph",
    reportNumber = "CERN-PH-TH-2013-251",
    doi = "10.1142/S0218301314500190",
    journal = "Int. J. Mod. Phys. E",
    volume = "23",
    number = "4",
    pages = "1450019",
    year = "2014"
}

@article{Bozek:2012en,
    author = "Bozek, Piotr and Broniowski, Wojciech",
    title = "{Charge conservation and the shape of the ridge of two-particle correlations in relativistic heavy-ion collisions}",
    eprint = "1204.3580",
    archivePrefix = "arXiv",
    primaryClass = "nucl-th",
    doi = "10.1103/PhysRevLett.109.062301",
    journal = "Phys. Rev. Lett.",
    volume = "109",
    pages = "062301",
    year = "2012"
}

@article{Braun-Munzinger:2023gsd,
    author = "Braun-Munzinger, Peter and Redlich, Krzysztof and Rustamov, Anar and Stachel, Johanna",
    title = "{The imprint of conservation laws on correlated particle production}",
    eprint = "2312.15534",
    archivePrefix = "arXiv",
    primaryClass = "nucl-th",
    doi = "10.1007/JHEP08(2024)113",
    journal = "JHEP",
    volume = "08",
    pages = "113",
    year = "2024"
}

@article{Vovchenko:2018fiy,
    author = {Vovchenko, Volodymyr and D\"onigus, Benjamin and Stoecker, Horst},
    title = "{Multiplicity dependence of light nuclei production at LHC energies in the canonical statistical model}",
    eprint = "1808.05245",
    archivePrefix = "arXiv",
    primaryClass = "hep-ph",
    doi = "10.1016/j.physletb.2018.08.041",
    journal = "Phys. Lett. B",
    volume = "785",
    pages = "171--174",
    year = "2018"
}

@article{Vovchenko:2019kes,
    author = {Vovchenko, Volodymyr and D\"onigus, Benjamin and Stoecker, Horst},
    title = "{Canonical statistical model analysis of p-p , p -Pb, and Pb-Pb collisions at energies available at the CERN Large Hadron Collider}",
    eprint = "1906.03145",
    archivePrefix = "arXiv",
    primaryClass = "hep-ph",
    doi = "10.1103/PhysRevC.100.054906",
    journal = "Phys. Rev. C",
    volume = "100",
    number = "5",
    pages = "054906",
    year = "2019"
}

@article{Vovchenko:2019pjl,
    author = "Vovchenko, Volodymyr and Stoecker, Horst",
    title = "{Thermal-FIST: A package for heavy-ion collisions and hadronic equation of state}",
    eprint = "1901.05249",
    archivePrefix = "arXiv",
    primaryClass = "nucl-th",
    doi = "10.1016/j.cpc.2019.06.024",
    journal = "Comput. Phys. Commun.",
    volume = "244",
    pages = "295--310",
    year = "2019"
}

@article{ALICE:2022xpf,
    author = "Acharya, Shreyasi and others",
    collaboration = "ALICE",
    title = "{Closing in on critical net-baryon fluctuations at LHC energies: Cumulants up to third order in Pb--Pb collisions}",
    eprint = "2206.03343",
    archivePrefix = "arXiv",
    primaryClass = "nucl-ex",
    reportNumber = "CERN-EP-2022-111",
    doi = "10.1016/j.physletb.2022.137545",
    journal = "Phys. Lett. B",
    volume = "844",
    pages = "137545",
    year = "2023"
}

@article{ALICE:2022xiu,
    author = "Acharya, Shreyasi and others",
    collaboration = "ALICE",
    title = "{First Measurement of Antideuteron Number Fluctuations at Energies Available at the Large Hadron Collider}",
    eprint = "2204.10166",
    archivePrefix = "arXiv",
    primaryClass = "nucl-ex",
    reportNumber = "CERN-EP-2022-040",
    doi = "10.1103/PhysRevLett.131.041901",
    journal = "Phys. Rev. Lett.",
    volume = "131",
    number = "4",
    pages = "041901",
    year = "2023"
}

@article{Speed1983,
    author = "Speed, T. P.",
    title = "{Cumulants and partition lattices}",
    journal = "Australian Journal of Statistics",
    volume = "25",
    number = "2",
    pages = "378--388",
    year = "1983",
    doi = "10.1111/j.1467-842X.1983.tb00391.x"
}

@article{ALICE:2024rnr,
    author = "Acharya, Shreyasi and others",
    collaboration = "ALICE",
    title = "{Probing Strangeness Hadronization with Event-by-Event Production of Multistrange Hadrons}",
    eprint = "2405.19890",
    archivePrefix = "arXiv",
    primaryClass = "nucl-ex",
    reportNumber = "CERN-EP-2024-151",
    doi = "10.1103/PhysRevLett.134.022303",
    journal = "Phys. Rev. Lett.",
    volume = "134",
    number = "2",
    pages = "022303",
    year = "2025"
}

@article{Savchuk:2021aog,
    author = "Savchuk, Oleh and Vovchenko, Volodymyr and Koch, Volker and Steinheimer, Jan and Stoecker, Horst",
    title = "{Constraining baryon annihilation in the hadronic phase of heavy-ion collisions via event-by-event fluctuations}",
    eprint = "2106.08239",
    archivePrefix = "arXiv",
    primaryClass = "hep-ph",
    doi = "10.1016/j.physletb.2022.136983",
    journal = "Phys. Lett. B",
    volume = "827",
    pages = "136983",
    year = "2022"
}

@article{Pradeep:2022eil,
    author = "Pradeep, Maneesha Sushama and Stephanov, Mikhail",
    title = "{Maximum Entropy Freeze-Out of Hydrodynamic Fluctuations}",
    eprint = "2211.09142",
    archivePrefix = "arXiv",
    primaryClass = "hep-ph",
    doi = "10.1103/PhysRevLett.130.162301",
    journal = "Phys. Rev. Lett.",
    volume = "130",
    number = "16",
    pages = "162301",
    year = "2023"
}

@article{Fokin:2025hbv,
    author = "Fokin, Ilya",
    title = "{Higher-order fluctuations: Unveiling the final frontier of QCD at the LHC with ALICE}",
    eprint = "2510.10847",
    archivePrefix = "arXiv",
    primaryClass = "nucl-ex",
    doi = "10.1051/epjconf/202636415003",
    journal = "EPJ Web Conf.",
    volume = "364",
    pages = "15003",
    year = "2026"
}

@article{Skokov:2012ds,
    author = "Skokov, V. and Friman, B. and Redlich, K.",
    title = "{Volume Fluctuations and Higher Order Cumulants of the Net Baryon Number}",
    eprint = "1205.4756",
    archivePrefix = "arXiv",
    primaryClass = "hep-ph",
    reportNumber = "BNL-98044-2012-JA",
    doi = "10.1103/PhysRevC.88.034911",
    journal = "Phys. Rev. C",
    volume = "88",
    pages = "034911",
    year = "2013"
}

@article{ALICE:2022wwr,
    author = "{ALICE Collaboration}",
    title = "{Letter of intent for ALICE 3: A next-generation heavy-ion experiment at the LHC}",
    eprint = "2211.02491",
    archivePrefix = "arXiv",
    primaryClass = "physics.ins-det",
    reportNumber = "CERN-LHCC-2022-009, LHCC-I-038",
    month = "11",
    year = "2022"
}

@techreport{Dainese:2925455,
      author        = "Dainese, A and Di Mauro, A",
      title         = "{Scoping document for the ALICE 3 detector}",
      institution   = "CERN",
      reportNumber  = "CERN-LHCC-2025-002, LHCC-G-185",
      address       = "Geneva",
      year          = "2025",
      url           = "https://cds.cern.ch/record/2925455",
}
\end{document}